\documentclass{SciPost}
\usepackage[T1]{fontenc}
\usepackage{mlmodern}
\usepackage{graphicx}
\usepackage{dcolumn}
\usepackage{bm}
\usepackage{amsmath}
\usepackage{amssymb}
\usepackage{slashed}
\usepackage{accents}
\usepackage{subcaption}
\usepackage[percent]{overpic}
\usepackage[bitstream-charter]{mathdesign}
\hypersetup{
    colorlinks,
    linkcolor={red!50!black},
    citecolor={blue!50!black},
    urlcolor={blue!80!black}
}
\DeclareSymbolFont{usualmathcal}{OMS}{cmsy}{m}{n}
\DeclareSymbolFontAlphabet{\mathcal}{usualmathcal}

\fancypagestyle{SPstyle}{
\fancyhf{}
\lhead{\colorbox{scipostblue}{\bf \color{white} ~SciPost Physics }}
\rhead{{\bf \color{scipostdeepblue} ~Submission }}

\fancyfoot[C]{\textbf{\thepage}}
}

\def\d{d\kern-.8 ex\vrule height 1.3 ex depth-1.24 ex width .7 ex \kern .15 ex}
\def\D{D\kern-1.7 ex\vrule height .87 ex depth-.8 ex width .7 ex \kern .95 ex}

\newcommand{\be}{\begin{equation}}
\newcommand{\ee}{\end{equation}}
\newcommand{\bea}{\begin{eqnarray}}
\newcommand{\eea}{\end{eqnarray}}

\begin{document} 

\pagestyle{SPstyle}

\begin{center}{\Large \textbf{\color{scipostdeepblue}{
Berry Picking: Random Wave Chaos Hierarchy for BPS Microstate Geometries
}}}\end{center}


\begin{center}\textbf{
Vladan {\D}uki\'c,\textsuperscript{1$\star$},
Milica Stepanovi\'c\textsuperscript{1,2$\dagger$} and
Mihailo \v{C}ubrovi\'c\textsuperscript{1$\ddagger$}
}\end{center}

\begin{center}
{\bf 1} Center for the Study of Complex Systems, Institute of Physics Belgrade, University of Belgrade, Pregrevica 118, 11080 Belgrade, Serbia
\\
{\bf 2} Faculty of Physics, University of Belgrade, Studentski Trg 12-16, 11000 Belgrade, Serbia
\\[\baselineskip]
$\star$ \href{mailto:email1}{\small djukic@ipb.ac.rs}\,\quad
$\dagger$ \href{mailto:email2}{\small mistepanovic@gmail.com}\,\quad
$\ddagger$ \href{mailto:email3}{\small mcubrovic@gmail.com}
\end{center}

\section*{\color{scipostdeepblue}{Abstract}}
\textbf{\boldmath{%
We estimate the strength of chaos of probe waves and probe geodesics in different smooth supergravity backgrounds of decreasing supersymmetry and/or increasing length of the AdS throat in the interior. Since BPS spectra are highly degenerate and thus do not display the level repulsion that ordinarily diagnoses chaos, we propose compliance with the Berry random wave hypothesis for probe wavefunctions as a new notion of BPS chaos suited to the supergravity regime. The wave chaos becomes stronger and stronger with less supersymmetry and longer throats (i.e., as we approach black hole solutions). Geodesic motion shows the opposite trend, becoming more and more regular. We explain this dichotomy by the existence of stable periodic orbits inside long throats while the overall measure of KAM tori decreases. Computing the Rényi entropies for the dual CFT states in the weak coupling regime, we show that they do not have such universal behavior.
}}

\vspace{10pt}
\noindent\rule{\textwidth}{1pt}
\tableofcontents
\noindent\rule{\textwidth}{1pt}
\vspace{10pt}

\section{Introduction}

Black holes are the prime theoretical labs for studying quantum gravity. Constructing the quantum state of a black hole at strong coupling and understanding its structure, dynamics and quantum information properties is pretty much the Holy Grail of modern quantum gravity research. On the other hand, black holes seem to be characterized by chaos. The chaotic dynamics of black hole states is among the crucial elements to understand their microscopic physics and quantum information properties (the Page curve and the firewall problem) \cite{Shenker:2013pqa,Polchinski:2015cea,Cotler:2016fpe,Saad:2019lba,Almheiri:2020cfm}. But when one tries to nail down the notion of ``black hole chaos`` to a specific, well-defined quantity to be computed, things become murky. Is the horizon the nest of chaos \cite{Shenker:2013pqa,Shenker:2014cwa}? Or the whole black hole wavefunction itself \cite{Balasubramanian:2022gmo}? Or the photon ring \cite{Giataganas:2024hil,Giataganas:2026ctn}? Another essential ingredient is the AdS/CFT correspondence (holography, gauge/gravity duality): it relates the semiclassical black holes in asymptotically anti-de Sitter spacetimes (spacetimes with negative cosmological constant) in (super)gravity to the strongly coupled, large-$N$ (large number of colors) gauge theories at or near the quantum critical/conformal invariant regime. These are known to be maximally chaotic (having the largest possible quantum Lyapunov exponent), a reflection of their black hole gravity duals. But there, again, it is not clear how exactly the \emph{black hole chaos}, i.e. the chaos in the ``bulk`` (AdS spacetime) relates to the ``boundary`` chaos, i.e. the chaos in the CFT.

So what is it that we do know for sure about black holes and chaos? It is the paradigm of fast scrambling \cite{Sekino:2008he}, out-of-time ordered correlators (OTOC) and maximum chaos bound \cite{Maldacena:2015waa}, and also the firm relation to random matrix theory (RMT) ensembles in two spacetime dimensions, described by Jackiw-Teitlboim (JT) gravity in AdS${}_2$ \cite{Saad:2019lba} which, however, may be special to 2D (but still relevant at least for near-extremal, i.e. low-temperature black holes, whose near-horizon geometry has a long AdS region, usually called ``AdS throat``). But as we said above, the maximum chaos of OTOC correlation functions is a statement about the \emph{holographic gauge theory at large number of colors} whereas the black hole/random matrix results of \cite{Cotler:2016fpe,Saad:2019lba} are statements about the \emph{bulk, i.e. the black hole itself.} We do not understand how exactly these two relate to each other in higher dimension, when we do not have just AdS${}_2$. But whatever happens in the CFT has to happen in the bulk, too. So there must be a ``holographic dictionary entry`` of chaos -- how the CFT notions translate to the bulk.

Here the subtle issue of classical vs. quantum chaos also enters the scene. We know that the strongly correlated, large-$N$ regime of the CFT corresponds to classical (super)gravity, therefore, in this regime, we should study the classical chaos in the bulk. Nevertheless, the typical probes are various fields that exhibit wave chaos, which is much closer to quantum chaos. But quantum chaos itself is only uniquely defined if there is a relevant semiclassical regime -- otherwise we just have various ``signatures of chaos`` \cite{Haake:book}, which in general tell us different things. This is part of the reason for so much confusion when it comes to ``black hole chaos`` -- \emph{there is no unique notion of quantum chaos away from the semiclassical limit.}

The kind of quantum chaos that is perhaps closest to a textbook piece, entering the cultural baggage of most physicists, is the Bohigas-Giannoni-Schmidt conjecture that quantum-chaotic Hamiltonians are well-described by Gaussian random matrices. The RMT then predicts the celebrated level repulsion and the Wigner-Dyson statistics of neighboring level spacings. A related quantity is the spectral form factor (SFF), which for quantum chaotic systems has the universal structure dip-ramp-plateau. Here, the ramp (linear growth) is the universal part, stemming from level repulsion, i.e. Wigner-Dyson statistics. The results of \cite{Cotler:2016fpe,Saad:2018bqo}, finding the linear ramp in the SFF for black holes in AdS${}_2$ gravity, are probably the most concrete proofs of black hole chaos that we have.

But on the other hand there is the more dynamical definition of chaos -- how the system behaves when perturbed. In this sense, the natural indicator is the Lyapunov exponent and its quantum counterpart, the quantum Lyapuov exponent of the OTOC. In classical gravity, the corresponding indicators are probe objects -- geodesics, fields, branes etc. -- moving in a given background. These do not probe the system in its given state but add a new object (Hamiltonian, equation of motion etc) whose behavior in the chosen background diagnoses something about this background.

It is far from clear what the relation is among different definitions and indicators of chaos. So far, various authors, including some of us, have studied holographic OTOC functions \cite{Shenker:2013pqa,Shenker:2014cwa}, bulk geodesics \cite{Basu:2011di,Djukic:2023dgk,Giataganas:2021ghs,Berenstein:2023vtd,Berenstein:2025ese}, energy spectrum \cite{Cotler:2016fpe,Saad:2019lba}, spectrum of BPS operators \cite{Chen:2024oqv} and the Berry curvature of states \cite{Chen:2026vml}. The last two approaches are specially tailored for supersymmetric systems, which cannot show textbook level repulsion as supersymmetric systems have huge energy degeneracy. This brings us to another puzzle that we want to address -- chaos in supersymmetric systems.

Supersymmetry is ubiquitous in string theory and controlled (``top-down``) constructions of black holes in string theory. Non-perturbative aspects of the theory that are captured by D-branes have been extremely fruitful in uncovering many aspects of black hole physics, one of them being the explicit counting of black hole microstates in the weak coupling regime of the parameters and matching it exactly to the entropy of a black hole in the supergravity regime \cite{Strominger:1996sh,Dijkgraaf:1996it,Maldacena:1997de}. But these crucially depend on supersymmetry. Supersymmetry leads to non-renormalization (protection) of some quantities, allowing us to compute them in weakly coupled CFT and translate to the strongly coupled CFT/classical supergravity regime.

However, black holes are not the only supersymmetric solutions. There are perfectly smooth solutions, with no horizons or singularities, which nevertheless look like a black hole to a distant observer, and to some extent reproduce some of black hole phenomenology. These solutions, so-called fuzzballs, are at the forefront of an ambitious program \cite{Bena:2013pda,Warner:2019jll,Bena:2022ldq,Bena:2025pcy,Raju:2018xue} aiming to resolve the black hole paradoxes by arguing that black holes in string theory do not exist -- instead, they should be understood as ensemble averages or smeared-out solutions of the fuzzball class, thus the horizons and singularities are just the artifacts of an effective theory. To what extent this portfolio is promising depends on the extent to which it can explain the known phenomena in black hole physics -- including chaos. A central goal of our present research is to understand if and how much of the black hole dynamics, both in the bulk and in CFT, can be reproduced by smooth solutions such as fuzzballs.



The fuzzball puzzle has been the subject of many studies, in particular on holographic covering. Holographic covering is a discussion on how one can classify BPS states into two distinct classes based on the following criterion: if we follow a given BPS state all the way from finite $N$ to the large $N$ regime, does it remain a BPS state throughout? States for which such a lift does exist are called \textit{monotonous}, whereas the ones for which it does not exist are called \textit{fortuitous}. There is a conjecture that, in the context of AdS/CFT, monotone BPS states will correspond to graviton gas in the bulk, and thus only fortuitous states might correspond to genuine black hole microstates \cite{Chang:2024zqi}. In other words, while fortuitous states can be anything (not necessarily black holes), monotone states definitely cannot be black holes. 

We thus again arrive at quantum chaos as a natural playground for testing these ideas. Here one encounters an additional difficulty: how can we ever have chaos in a state with large degeneracy, such as a BPS state? Degeneracy certainly precludes any notion of chaos in the sense of RMT statistics of opertors. This problem was addressed in the ``LMRS`` series of papers \cite{Lin:2022zxd, Lin:2022rzw}, where the idea is to project the operators to a certain subspace, which breaks degeneracy. Within this new notion of BPS chaos, strong indications were found that chaos is closely related to the existence of fortuitous BPS states \cite{Chen:2024oqv}. Since these are the states that are not BPS at large $N$, they will be exploring the non-BPS sector in the spectrum and upon tuning $N$ down will carry possible chaotic signatures back into the BPS part of the spectrum. The final outcome of \cite{Chen:2024oqv} is a hierarchy of increasing chaos with decreasing amount of supersymmetry, i.e. the number of supercharges. Still, no fuzzball states studied in \cite{Chen:2024oqv} come anywhere near the strong chaos expected from black holes. The amount of chaos (strong or weak) is measured in this context by Thouless time, the time needed to enter the linear ramp regime in the SFF. If it scales as $N$ to some positive power it will become infinite in the large-$N$ limit, at which point it will thus correspond to the absence of chaos. If it is an order-unity number, it will correspond to strong chaos. The recent important followup \cite{Chen:2026vml}, which goes beyond the LMRS perspective for operators and studies the chaos in \emph{states}, again results in the same conclusion: monotone -- no chaos, fortuitous -- RMT behavior.

Previous studies of chaos in these contexts have been done at finite $N$ and at vanishing t’ Hooft coupling $\lambda \equiv g_s N \ll 1$. Since we have explicit holographic gravitational duals of these states at large $N$ and $\lambda \gg 1$, the natural task is to do a similar computation in this regime. Unfortunately, the energy-spectrum-related notions of chaos (level spacing, SFF) are very hard to obtain in gravity as we do not know the full Hilbert space of semiclassical supergravity solutions. This motivates us to explore other notions of chaos in BPS contexts.

First of all, we have to give up the study of eigenvalue chaos and instead focus on the eigenvector chaos. Just as the level repulsion and the Wigner statistics are the signs of RMT chaos in the eigenvalues, the consequence of RMT for the structure of the states in encapsulated in the Berry random wave conjecture. This is our proposal for a new notion of BPS chaos in the supergravity regime.

In summary, the main questions we want to answer are the following:
\begin{enumerate}
\item What are the bulk indicators of black hole chaos in the classical supergravity regime? Is it the Berry random wave behavior?
\item What is the correct notion of chaos for BPS (supersymetric) systems from the bulk viewpoint?
\item Is there a systematic trend in chaotic behavior as we progressively reduce supersymmetry and approach the black hole behavior?
\end{enumerate}

\subsection{The backgrounds studied}

The goal is now to study an array of progressively less and less supersymmetric backgrounds in supergravity.\footnote{We are obviously inspired by \cite{Chen:2024oqv,Chen:2026vml} to employ such strategy.} To this end, we will study the 1/2-BPS Lin-Lunin-Maldacena (LLM) bubbling geometries, 1/4-BPS 2-charge supertubes and 1/8-BPS 3-charge supertubes (closely related to the Lunin-Maldacena (LM) solutions) and the 1/8-BPS superstratum geometries.

In string theory, we know that at 1/2-BPS there is not enough entropy to form a black hole. Still, the bubbling solutions of Lin-Lunin-Maldacena can serve as a useful toy model for black hole physics: for example, they can exhibit long-lived trapping and are useful for understanding the role of averaging in gravity \cite{Berenstein:2023vtd, Berenstein:2025ese}. These geometries correspond to the 1/2-BPS sector of $\mathcal N=4$ super-Yang-Mills theory and are all monotone.

At 1/4-BPS we study a 2-charge supertube in the D1-D5 theory (in type IIB frame; equivalently, F1-P in the heterotic frame, or NS5-P in type IIA), which has a vanishing horizon area and thus cannot reproduce the black hole entropy, but higher derivative corrections give it a finite entropy that scales as $\sim\left(N_1N_P\right)^{1/2}$ (in the IIA frame) \cite{Dabholkar:2004yr}. 

To get a black hole with a macroscopic horizon, we need to add NS5 branes into the game to obtain an entropy that scales as $\sim\left(N_1N_5N_P\right)^{1/2}$. In the IIB frame where one is dealing with the D1-D5-P system, examples are the 3-charge supertubes and a huge family of solutions called superstrata (for a review see \cite{Shigemori:2020yuo} and references therein). Since these solutions have been built out of CFT states that are quantized in units of $\sim 1/R_y$ (where $R_y$ is the compactification radius of the momentum charge P), instead of $\sim 1/N_1N_5 R_y$, they give a ``sub-sub-leading`` contribution to the entropy $\sim \left(N_1 N_5\right)^{1/2} \left(N_P\right)^{1/4}$ \cite{Shigemori:2019orj,Mayerson:2020acj}. They thus correspond to supergraviton gas in the bulk and are duals of monotonous BPS states; nonetheless, these solutions still show some features of typical black hole microstates, such as having a mass gap in the spectrum of order $\sim 1/N_1N_5R_y$ \cite{Bena:2018bbd}. These are a priori expected to be the ``most black-holish`` systems that we study.\footnote{Here we will work solely with monotonous states in the classification of the holographic covering. There exists a family of superstrata solutions that are dual to states with fractionated momentum \cite{Bena:2016agb}, which are a natural candidates for a fortuitous states. We will leave the study of supergravity duals of fortuitous states for the future \cite{Bena:2025pcy,Hughes:2025tdy}. }

The picture so far is that there is a clear gradation of progressively stronger chaos as the supersymmetry is lowered. On one hand, this is expected as proper, finite-horizon-area black holes can only be 1/16-BPS in super-Yang-Mills or 1/8-BPS in the D1-D5 type IIB framework, ``small black holes`` with vanishing horizon can be at most 1/4-BPS, and at 1/2-BPS we only get ``incipient black holes``, i.e. naked singularities a la Gubser that are "good" and can be enclosed by a horizon after a suitable deformation. On the other hand, we can clearly have non-black-hole states even at 1/16-BPS or indeed with no supersymmetry at all. So the question is: do ``BPS chaos`` indicators detect black-holishness, or merely the breaking of SUSY?

\subsection{Outline of the results}

Our strategy is to study the chaotic properties of probe scalar waves and probe geodesics for the  collection of examples listed above. We compute the solutions of the massless wave equation in the coordinate representation for the 1/2-BPS LLM geometries, for the 1/4-BPS D1-D5 supetubes, and for 1/8-BPS D1-D5-P supertubes and superstrata. The computation is done with Dirichlet boundary conditions that correspond to looking at a specific CFT state (without deforming the theory).

We will inspect the strength of chaos by comparing the statistical properties of the solutions with the Berry random wave conjecture, a consequence of RMT which states that chaotic eigenfunctions are well-approximated by sums of many waves with fixed energy and random phases. We will uniformly find that black-holishness of the background increases chaos -- reducing the supersymmetry, increasing the length of the AdS throat (which is infinite in extremal black holes) and developing a (naked) singularity all make the probe waves more chaotic. In this sense, it seems the the CFT chaos in terms of OTOC exponents (known to reach its maximum in black hole backgrounds) and the RMT description of AdS${}_2$ throats neatly go hand in hand with the bulk probes of chaos. Importantly, increasing chaos is \emph{not} merely about breaking supersymmetry, as it also increases with the lengthening of the throat or developing singularities -- properties with no direct relation to the number of supercharges.

The second important finding is the behavior of bulk \emph{geodesics} (as opposed to waves). It is a long-standing conundrum that, on the one hand, semiclassical black hole horizons give rise to fast scrambling and maximum chaos in holographic CFTs, and on the other hand most black holes yield separable Hamilton-Jacobi equations and consequently integrable dynamics for probe geodesics \cite{Chervonyi:2013eja}. The converse is also true: we tend to think of horizonless and smooth geometries as being less chaotic than black holes, yet they quite often (though not always) have nonintegrable geodesic dynamics \cite{Chervonyi:2013eja,Bena:2017upb,Bianchi:2020des,Berenstein:2023vtd,Berenstein:2025ese}. And indeed, in this paper we find the same: geodesics become \emph{less} chaotic as the waves become \emph{more} chaotic. They are most chaotic in LLM geometries, and least so in long-throat superstrata.

This counterintuitive result is shown to be rooted in the specifics of wave dynamics: true, wave chaos increases with decreasing supersymmetry and increasing probe lengths, and in the classical limit this means the measure of KAM tori decreases, giving way to the chaotic sea, but at the same time the number of stable periodic orbits grows. Their spectrum is discrete, and their measure is thus zero, so they do not influence the global properties of waves, but they can significantly influence the dynamics of geodesics, which are local objects. In other words, the orbits spend most of their time inside the long throat, where quasi-integrals of motion (approximate symmetries) decrease the chaos (the longer the throat, the more pronounced the effect). The waves are everywhere at the same time and are not so sensitive to these throat-only approximate symmetries.

Finally, we have connected our bulk results to the complexity of CFT states benchmarked by the Shannon and participation entropy (both being special cases of the Renyi entropy). Although direct comparison is not feasible since the tractable, weakly coupled CFT regime does not coincide with the strongly coupled supergravity regime, we can at least put the lower bound on complexity: the weak coupling result can likely only grow larger, not smaller, when going to strong coupling. Here, we the BPS hierarchy is slightly changed: LLM backgrounds have zero Shannon entropy, 2-charge supertubes and superstrata both have Shannon entropy scaling as $\log\sqrt{N}$, but 3-charge supertubes have Shannon entropy of order $\log N$, just like black holes!\footnote{Notice that these entropies are entirely different from the thermodynamic, Bekenstein-Hawking entropy.} This is simply because 3-charge supertubes have many different quantum numbers while superstrata are defined by at most 3 quantum numbers. Also, the Shannon entropy is insensitive to the throat length. In other words, the BPS hierarchies in supergravity is simpler and sharper, while on the CFT side it depends on the details of the state (if we characterize it through Renyi entropies).

The ultimate goal -- understanding the microscopic properties of black hole chaos -- remains elusive for now. Since the formation of a horizon is a non-perturbative phenomenon, it is not clear if we can directly extrapolate our results from smooth solutions to black holes. This remains our goal for further work.

The structure of the paper is as follows. In the next section we explain the Berry random wave paradigm, the main formalism for diagnosing the chaos of bulk probes. Section 3 deals with waves and geodesics in LLM geometries, and section 4 with the same kind of probes for supertubes and superstrata. Section 5 brings the Shannon and participation entropy computations for these solutions, and some general implications of the results found. Section 6 sums up the conclusions.

\section{Quantum chaos, eigenstates and the Berry conjecture}

We expect that both energy levels and the corresponding eigenstates of a quantum chaotic system should know about chaos. The result for the levels is famous: according to the Wigner surmise, the Hamiltonian of a quantum chaotic system in a generic basis is well-described by an element of a Gaussian ensemble of random matrices (from the orthogonal, symplectic or unitary ensemble, depending on the symmetry class), and consequently the energy spacings $s_i\equiv E_{i+1}-E_i$ follow the Wigner-Dyson probability distribution function:
\be
P(s)=\mathcal{N}_\alpha s^{\alpha-1}e^{-A_\alpha s^2},\label{wd}
\ee
where $\alpha=2,3,5$ for GOE, GUE or GSE respectively. Since $P(0)=0$, the levels show repulsion and never cross. While the Wigner surmise was never rigorously proven and goes under the name of the Bohigas-Giannoni-Schmidt (BGS) conjecture, it is very well supported by examples and can be taken to be more or less universally valid -- unless, of course, some of the assumptions fail. One such assumption is the absence of any accidental symmetries that correlate different energy levels. Any amount of supersymmetry is precisely such a symmetry -- it introduces large degenerate sectors related by the action of supercharges. In BPS systems, the spectrum disintegrates into discrete spikes containing multiple states, and the BGS conjecture fails. The LMRS criterion of BPS chaos, proposed in \cite{Lin:2022zxd} and put to use in \cite{Chen:2024oqv}, is designed precisely to overcome this difficulty. It comes as close as possible (or as close anybody has so far made it) to the direct generalization of the Wigner surmise for supersymmetric systems.

But the states also know about chaos. Following similar logic as for the Gaussian statistics of the spectrum, Michael Berry has conjectured (originally mainly for quantum billiards) that the wavefunction of a quantum chaotic state $\Psi_n$ (i.e., the state vector represented in some basis), for $n$ sufficiently large, is well-described by a superposition of a large number $L$ of \emph{monochromatic} waves (i.e., waves with fixed wavenumber $k_n$) with \emph{uniform equidistant} wavevectors and \emph{uniform random} phases $\phi_j$ ($j=1\ldots L$) \cite{Berry:1977}:
\be
\Psi_n(\mathbf{x})\leftrightarrow\lim_{L\to\infty}\sum_{j=1}^L\exp\left[\imath k_n\left(x\cos\left(\frac{2\pi}{L}j\right)+y\sin\left(\frac{2\pi}{L}j\right)+\phi_j\right)\right],\label{eq:berrycon}
\ee
where $\mathbf{x}=(x,y)$. The above formulation is clearly for a two-dimensional system but higher-dimensional generalization is straightforward: we have instead a $D$-dimensional lattice of uniform equidistant wavevectors.

As we have stated, the original claim was made mainly to explain the chaotic properties of quantum billiards, where the Hamiltonian consists of only the kinetic term and the wave equation is just the Helmholtz equation inside the billiard. However, the reasoning of the original paper \cite{Berry:1977} and also of later developments \cite{Berry:1981mom,Berry:1986lxu,Berry::LesHouches,Urbina:2007fuq} is independent of the exact form of the potential, which in principle can be arbitrary. The potential enters eq.~(\ref{eq:berrycon}) through the wavenumber $k_n$, which, of course, depends on the total energy. As long as this is taken into account, the hypothesis remains viable provided the initial assumption of strong mixing (at least locally), leading to universality, is met. Experiments with microwave billiards have decisively supported the hypothesis \cite{Stockmann:book}, but it is also very well supported by experiments in other systems, e.g. mesoscopic systems like metal grains \cite{Mirlin:2000,Mirlin:2002}.

Once we accept the conjecture in the form (\ref{eq:berrycon}), we can compute the moments and correlation functions of the system. It turns out that the two-point correlator has a simple and universal form, in terms of Bessel functions, which precisely corresponds to \emph{Gaussian random fields}. This is mathematically nontrivial (although perhaps intuitively expected): uniform statistics of random \emph{phases} leads to a function whose \emph{values} come from a Gaussian random field. Indeed, one can invert the logic and start from the assumption that a strongly chaotic wavefunction behaves as a Gaussian random field, and then derive the statistical equivalence with the random wave sum of eq.~(\ref{eq:berrycon}). For details on how various subtly different formulations of the hypothesis relate to each other, see \cite{Urbina:2003xip,Urbina:2007fuq}. The correlation function, defined as
\be
C(\mathbf{x}_1,\mathbf{x}_2)\equiv\langle\Psi_n(\mathbf{x}_1)\Psi_n(\mathbf{x}_2)\rangle\label{eq:corrfun}
\ee
for a system obeying the Berry random wave hypothesis reads (computed already in \cite{Berry:1977}):
\be
C_\mathrm{BRW}(\mathbf{x}_1-\mathbf{x}_2)=\frac{1}{\Gamma\left(\frac{D}{2}\right)}\left(\frac{2}{k_n\vert\mathbf{x}_1-\mathbf{x}_2\vert}\right)^{\frac{D-2}{2}}J_{\frac{D-2}{2}}\left(k_n\vert\mathbf{x}_1-\mathbf{x}_2\vert\right).\label{eq:berryconbessel}
\ee
The correlator (\ref{eq:berryconbessel}) is in practice the main indicator to test, much more convenient that generating a random field and comparing with it. Notice a few important features:
\begin{enumerate}
\item The correlator depends only on the difference $\mathbf{x}_1-\mathbf{x}_2$, not on the individual points. This can only be exactly satisfied if the whole phase space is chaotic with no stability islands and similar structures. Such a situation is rare in practice, but the correlator will still have very weak dependence on $\mathbf{x}_1$ and $\mathbf{x}_2$ separately as long as the chaotic component is large enough.
\item The correlator depends only on the modulus of the difference, not on its direction. Again, this will never by exactly true for most systems, but can be approximately true.
\item The wavenumber $k_n$ is assumed constant. This is the main limitation of the original conjecture, but not a necessary one, as we have mentioned above. 
\end{enumerate}
The last point is most often strongly violated in practice. Indeed, for a Hamiltonian of the form $H=\nabla_\mathbf{x}^2+V(\mathbf{x})$ whenever the potential is nontrivial, the wavenumber will be replaced by an $\mathbf{x}$-dependent function. Typical ways out are the following:
\begin{enumerate}
\item For a Klein-Gordon equation of the form $-\partial_t^2+\nabla_\mathbf{x}^2+V(\mathbf{x})=0$ we can assume a sort of ``microscopic chaos`` a la Boltzmann, where at least on small scales (though not necessarily on large scales) the mixing is strong enough that the potential can be approximated by its mean value within a region of interest (not necessarily the whole space but some region where the wavefunction is mainly supported):
\be
k_n^2=E_n^2-\bar{V}.\label{eq:berryknmean}
\ee
\item For the same Klein-Gordon equation, we can take into account some spatial dependence of the wavenumber by taking the mean value of the potential between the two points $\mathbf{x}_1$ and $\mathbf{x}_2$, rather than uniformly in a whole region of space:
\be
k_n^2\mapsto k_n^2(\mathbf{x})=E_n^2-V(\mathbf{x}).\label{eq:berrykn}
\ee
In this case, $C(\mathbf{x}_1,\mathbf{x}_2)$ from (\ref{eq:corrfun}) depends on both points separately, and on the right-hand side we can take $k_n\mapsto k_n\left(\frac{\mathbf{x}_2-\mathbf{x}_1}{2}\right)$. This clearly only makes sense when $\Delta\mathbf{x}$ is not too large, and the potential has no large gradients. Otherwise, it offers no benefit with respect to (\ref{eq:berryconbessel}). In fact, both ways out are hardly viable if the variation of the potential is very large.
\end{enumerate}

In our work, the potential will always be nontrivial but smooth, without large gradients and without singularities (except for one case). Also, since all our spacetimes will be asymptotically AdS, the effective potential will always be confining at large $r$ (i.e. at large $\vert\mathbf{x}\vert$), so the wavefunction will always be confined in real space. Therefore, our systems do not differ that much from a billiard, except that the constant potential is replaced by a nontrivial one. We will most often use the mean wavenumber of eq.~(\ref{eq:berryknmean}) to obtain the function (\ref{eq:berryconbessel}), but we will check that the other choice (\ref{eq:berrykn}) does not lead to qualitatively different conclusions.
\\

All of the above holds when the system is large enough that near-boundary effects are negligible. If this is not so, or if the boundary conditions are more complicated than the usual Dirichlet conditions, the statistics might change substantially \cite{Srednicki:1998,Urbina:2007fuq}.

\subsection{Porter-Thomas distribution}

When testing our data for the Berry-random-wave-type correlations, we will see that some systems give a good (or even excellent) agreement with the correlation function (\ref{eq:berryconbessel}) even though some other indicators suggest that chaos is weak. In such cases, one should take the agreement with the random-wave correlation function with a grain of salt, as it is an indicator that can yield false positives, if the classical limit, while near-integrable, exhibits KAM tori with sufficiently complicated structure.\footnote{The mechanism in this case is essentially that the two-point function is determined by the projection of the phase space to a two-dimensional manifold. Tori with sufficiently complicated metric will have complicated projections, that are well-approximated by a random field.

We thank Juan-Diego Urbina for pointing out this phenomenon to us.\label{footkam}} A more refined criterion in such cases is the existence of caustics in the wavefunction, typical for integrable and near-integrable systems. The crucial role of caustics in this case is predicted already in the first Berry hypothesis paper \cite{Berry:1977}.

The presence of caustics can be tested by studying the distribution of the wave intensity $I(\mathbf{r})\equiv\vert\psi(\mathbf{r})\vert^2$ throughout the spatial range of the solution (it is understood that the wavefunction is normalized to unity). The distribution function of intensity $P(I)$ is then predicted to obey the Porter-Thomas distribution for a quantum chaotic system in the Berry hypothesis regime. This is a simple consequence of assuming the wavefunction $\Psi_n$ to be a Gaussian random field and performing a change of variables $\vert\Psi\vert^2\mapsto I$. For a GOE ensemble, appropriate in our case, the outcome reads:
\be
P_\mathrm{TP}(I)=\frac{1}{\sqrt{2\pi I}}e^{-\frac{I}{2}}.\label{eq:porterthomas}
\ee
Notice that the distribution is normalized to unity even though it diverges at zero. Informally speaking, the intensity is mostly low and high-intensity spots are exponentially rare. In weakly chaotic or nonchaotic systems, the exponentially suppressed tail will show occasional spikes corresponding to high-intensity lines (caustics). These are rare events and thus the overall shape of the curve is still similar, but the detection of spikes is easier than the detection of small uniform deviations from the Bessel oscillations.




\section{Weak wave chaos in 1/2-BPS LLM systems}

Following \cite{Chen:2024oqv,Chen:2026vml}, we will progressively go from higher to lower supersymmetry but we will be careful to distinguish between the cases with the same amount of supersymmetry which nevertheless might differ in their ``proximity`` to black holes. We will start from the LLM geometries \cite{Lin:2004nb}, a well-known supergravity description of the 1/2-BPS sector in super-Yang-Mills. Let us first briefly summarize the LLM solution in supergravity and in the dual CFT, and then we will probe the chaos in the bulk (in the supergravity regime).

\subsection{Bubbling geometries in supergravity and in $\mathcal{N}=4$ super-Yang-Mills}

We will not repeat here the rather intricate and complex story of how the bubbling geometries in AdS come to be. For this, the reader can consult classic references such as \cite{Lin:2004nb,Berenstein:2004kk,Myers:2001aq,Balasubramanian:2005mg}. In short, starting from pure AdS${}_5\times\mathbb{S}^5$, dual to $\mathcal{N}=4$ super-Yang-Mills theory, we can introduce giant and antigiant gravitons. When a large number of them condense, they produce objects of dimension $\sim N^2$ which backreact on geometry. Such solutions keep 16 supercharges, so they are 1/2-BPS.

There are many different ways to encode the LLM sector from the CFT point of view. In the 1/2-BPS sector only two out of six super-Yang-Mills scalars are excited: $\Phi_1,\Phi_2\neq 0$ while $\Phi_3=\Phi_4=\Phi_5=\Phi_6=0$. They are combined into $Z=\Phi_1+\imath\Phi_2$ and its conjugate $\bar{Z}$. Chiral primaries are single-trace operators $\mathrm{Tr}(Z^k)$. Any operator can be expressed in terms of single-trace operators and their products, so one possible basis are the operators $\left(\mathrm{Tr}\left(Z\right)^k\right)^p$. This basis is overcomplete because of the trace relations. A general operator is a Z-word with an arbitrary number of traces inserted in every term.

In the supergravity regime ($N\gg 1$, i.e. $c\gg 1$) we can classify operators into (i) small, with $\Delta\sim 1$ (ii) giant, with $\Delta\sim N$ (iii) huge, with $\Delta\sim N^2$. The small and the giant ones are probes (fields or extended objects like branes), the huge ones backreact on the vacuum, i.e. geometry. Our analysis mainly deals with probes, both small and giant. We will use scalar probe fields in the bulk, which correspond to light operators with $\Delta=5/2$, and probe geodesics.

The geodesic limit requires some care. Since we are in the supergravity regime we certainly need large central charge: $c\sim N^2\gg 1$. A geodesic is obtained as the infinite-energy limit of fields (i.e., the eikonal limit), which means $\Delta\gg 1$, but in order to stay in the probe limit $\Delta$ cannot become comparable with $c$, so the regime is
\be
N^2\sim c\gg 1,~~\Delta\gg 1,~~\Delta/N\sim\mathrm{const.}
\ee
We will not attempt to do precision holography here, i.e. to identify the exact CFT operator that corresponds to our geodesic probes (although it remains a very intriguing task for later).

The final detail needed to understand the appeal of LLM geometries is the dual free fermion model found in \cite{Berenstein:2004kk} and studied in \cite{Takayama:2005yq,Mandal:2005wv}. The model consists of $N$ 1D fermions in harmonic potential in the first quantization \cite{Takayama:2005yq}:
\be
H=\sum_{j=1}^N\left(-\partial_i^2+\frac{1}{2}x^2\right).\label{eq:llmfermi}
\ee
The first-quantized creation and annihilation operators $c_i^\dagger$, $c_i$ bring the $i$-th fermion up or down in energy by one. The operator-state correspondence works as
\be
\mathrm{Tr}\left(Z^k\right)\Leftrightarrow\sum_{j=1}^N\sqrt{\frac{(j+k)!}{j!}}c^\dagger_{j+k}c_j\vert j\rangle.\label{eq:fermisimple}
\ee
The conclusion is now that any curve in the plane, defining the Fermi surface of the fermionic model (or, equivalently, dividing the plane into particles and holes, conventionally called black and white regions) defines an LLM solution. For example, a circular Fermi surface, i.e. a black disk (which corresponds to AdS${}_5\times\mathbb{S}^5$) is obtained as
\be
\vert\mathrm{disk}\rangle=\prod_{k=1}^Nc_k^\dagger\vert 0\rangle,~~E=\sum_{k=1}^N\frac{2k-1}{2}=\frac{N^2}{2}.\label{eq:fermifs}
\ee
For our purposes, we are mainly interested in disk+ring(s) configurations in the LLM plane. A black ring of outer radius $R'$ and inner radius $R$ can be obtained by cutting out a white disk of radius $R$ from a black disk of radius $R'$. In fermionic terms it means
\be
\vert\mathrm{ring}\rangle=\prod_{k=1}^Rc_k\prod_{l=1}^{R'}c_l^\dagger\vert 0\rangle,~~E=\frac{R'^2-R^2}{2}.\label{eq:fermiring}
\ee
The disk+ring configuration is just the superposition of (\ref{eq:fermifs}) and (\ref{eq:fermiring}). Denoting the disk radius by $R_1$ and the inner and outer radius of the ring by $R_2$ and $R_3$ respectively, we get:
\be
\vert\mathrm{disk+ring}\rangle=\prod_{j=1}^{R_1}c_j^\dagger\prod_{k=1}^{R_2}c_k\prod_{l=1}^{R_3}c_l^\dagger\vert 0\rangle,~~E=\frac{R_1^2+R_3^2-R_2^2}{2}.\label{eq:fermidiskring}
\ee
We can continue adding rings in this way, getting a disk+multiring configuration. In the basis of single-trace Z-words, the rings correspond to the Schur polynomials of single-trace operators \cite{Lin:2004nb}, and in supergravity they yield a spacetime which is asymptotically AdS${}_5\times\mathbb{S}^5$ but with only radial and time transition symmetry, resulting in non-separable equations of motion.

Now that we have understood how the CFT state defines the background, we can write the solution for the metric in type IIB supergravity. The ten-dimensional metric consists of:
\begin{enumerate}
\item A four-dimensional manifold $M_4$ with time $t$, a two-dimensional $(x_1,x_2)$ plane (the ``LLM plane`` where the black-and-white pattern is defined) and the orthogonal, off-plane coordinate $\xi$. 
\item The first three-sphere $\mathbb{S}^3$ defined by the flux of the five-form field $F_5$ through the LLM plane, whose radius shrinks to zero in the black regions.
\item The second three-sphere $\widetilde{\mathbb{S}^3}$, defined by the flux of the dual five-form field $\widetilde{F}_5$ through the LLM plane, whose radius shrinks to zero in the white regions.
\end{enumerate}
Explicitly, the metric reads \cite{Lin:2004nb}:
\bea
&&ds^2=-h^{-2}\left(dt+V_idx^i\right)^2+h^2\left(d\xi^2+dx_idx^i\right)+\xi e^Gd\Omega_3^2+\xi e^{-G}d\widetilde{\Omega}_3^2\nonumber\\
&&h^2=\frac{1}{\xi}\sqrt{\frac{1}{4}-z^2}, \quad \partial_{\xi} V_i=\frac{\epsilon_{ij} \partial_jz}{\xi},\quad z=\frac{1}{2}\tanh G.\label{eq:llmmetric}
\eea
We thus have a whole family of solutions determined by the function $z(\xi,x_1,x_2)$. The equation for $z$ is first-order as it follows from the Killing spinor equations that must be satisfied to preserve the supersymmetries:
\begin{equation}
\partial_i\partial_iz+\xi\partial_\xi\left(\frac{\partial_\xi z}{\xi}\right)=0.\label{eq:llmz}
\end{equation}
The line elements of two three-spheres (parametrized by $(\psi_1,\psi_2,\psi_3)$ and $(\tilde{\psi}_1,\tilde{\psi}_2,\tilde{\psi}_3)$ respectively) are given by $d\Omega_3$ and $d\widetilde{\Omega}_3$. For calculations we will mainly use a slightly different coordinate system for the four-manifold $M_4$, where the Cartesian coordinates $(x_1,x_2,\xi)$ are replaced by spherical ones $(r,\theta,\phi)$ with $0\leq\phi<2\pi$ the azimuthal angle and $-\pi/2<\theta\leq\pi/2$ the polar angle:
\bea
&&ds^2=-h^{-2}\left(dt+V_\phi d\phi\right)^2+h^2\left[r^2\left(\cos^2\theta d\phi^2+d\theta^2\right)+dr^2\right]+r\sin\theta \left(e^Gd\Omega_3^2+e^{-G}d\widetilde{\Omega}_3^2\right)\nonumber\\
&&V_\phi=r\cos\theta\left(V_2\cos\phi-V_1\sin\phi\right).\label{eq:llmmetricsph}
\eea
The relations between $h$, $z$ and $G$ stay the same as in eq.~(\ref{eq:llmmetric}) except for the obvious transformation $\xi\mapsto r\sin\theta$. It can be shown that black and white regions correspond to the regions in the LLM plane where $z(\xi=0,x_1,x_2)=+1/2$ and $z(\xi=0,x_1,x_2)=-1/2$ respectively.

The final ingredient is the existence of grayscale solutions. In the fermionic model, they obviously correspond to mixed states, obtained by coarse-graining black-and-white solutions. More generally, they can also be understood as coming from integrating out fluctuations to get a supergravity background where a state with long-range correlations looks mixed in the semiclassical regime \cite{Balasubramanian:2005mg,Balasubramanian:2018yjq,Berenstein:2025ese}. In this case, the function $z$ takes intermediate values in the LLM plane: $-1/2<z(\xi=0,x_1,x_2)<1/2$ and the scalar curvature diverges in the LLM plane. Nevertheless, it is a ``good`` singularity in the Gubser sense, often understood as an incipient black hole (with horizon radius equal to zero, but which will become finite when quantum corrections are taken into account). This is as close as we can get to a black hole at 1/2-BPS. 

\begin{figure}[ht]
\centering
(A)\includegraphics[width=.4\linewidth]{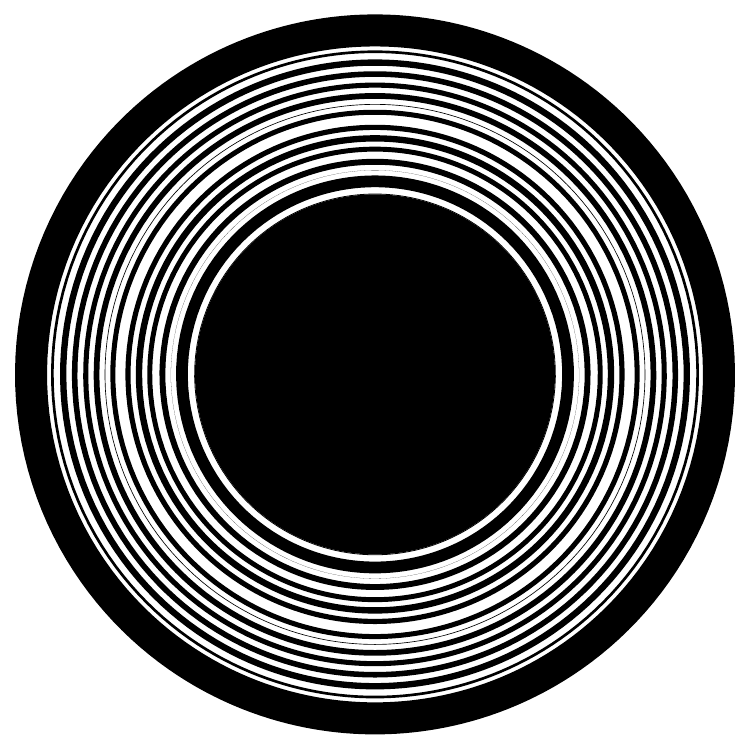}
(B)\includegraphics[width=.4\linewidth]{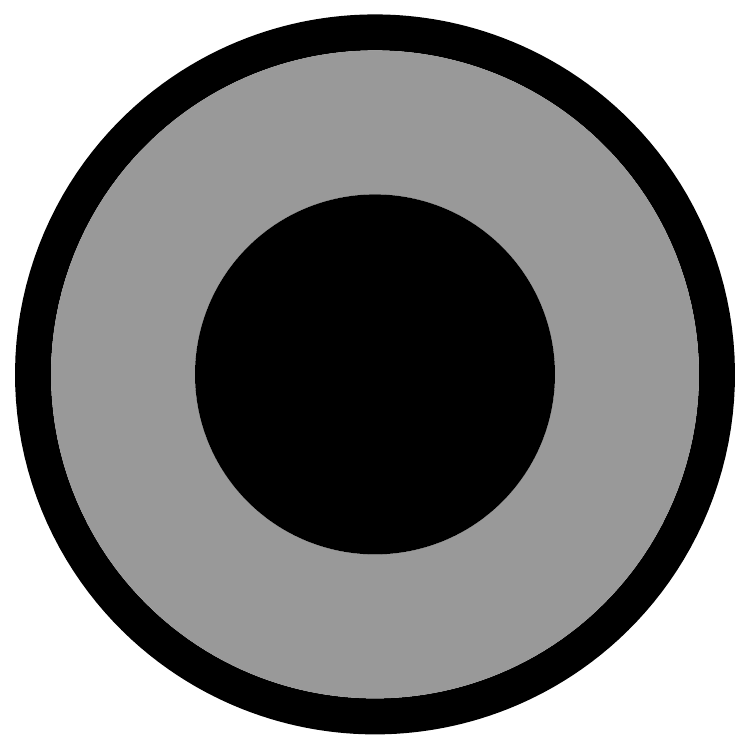}
\caption{The flux density (``nuance``) in the LLM plane for the disk+rings configuration (A) and the disk+gray ring configuration (B). The latter can obviously be thought of as a coarse-grained version of the former.}
\label{fig:llmbwgray}
\end{figure}

The two backgrounds that we test -- disk + rings and disk + gray rings -- are shown in Fig.~\ref{fig:llmbwgray}. The coarse-graining logic is now obvious: many thin black rings become a gray region in the coarse-grained picture.\footnote{While the details of the averaging mechanism are subtle and highly interesting, they are not necessary for our current story. They can be found, e.g., in \cite{Balasubramanian:2005mg,Balasubramanian:2018yjq,Berenstein:2025ese}.}

\subsection{Massless Klein-Gordon equation}

Let us now do our main task -- look at the dynamics of bulk probes. In this subsection we study the massless scalar wave (we believe a massive scalar would yield similar results). A massless scalar probe in the general LLM background has three groups of conserved angular momentum quantum numbers. In the $(t,r,\theta,\phi)$ manifold the conserved component is the projection of the angular momentum that we denote by $\ell$. On the sphere $\mathbb{S}^3$ we have the quantum numbers $k,l,m$ and on $\widetilde{\mathbb{S}^3}$ likewise $\tilde{k},\tilde{l},\widetilde{m}$. We adopting ansatz
\be
\Phi=e^{-\imath\omega t+\imath\ell\phi+\imath m\psi_1+\imath\widetilde{m}\tilde{\psi}_1}Y_{klm}(\psi_1,\psi_2,\psi_3)Y_{\tilde{k}\tilde{l}\widetilde{m}}(\tilde{\psi}_1,\tilde{\psi}_2,\tilde{\psi}_3)\phi(r,\theta),\label{eq:llmansatz}
\ee
where $Y$ are three-spherical harmonics. The Klein-Gordon equation now reads:
\bea
&&\partial_r^2\phi+\frac{\partial_\theta^2\phi}{r^2}+\frac{6\partial_\theta h\left(1-4z^2\right)+h\left(12z\partial_\theta z+\tan\theta\left(1-4z^2\right)\right)}{r^2h\left(4z^2-1\right)}\partial_\theta\phi\nonumber\\&&+\frac{\left(6r\partial_rh\left(1-4z^2\right)+2h\left(6rz\partial_rz+4z^2-1\right)\right)}{rh\left(4z^2-1\right)}\partial_r\phi\nonumber\\
&&+\left[\frac{2k_1\left(k_1+2\right)h^4}{1+2z}+\frac{2k_2\left(k_2+2\right)h^4}{1-2z}+\frac{\left(r^2\omega^2h^4-\sec^2\theta(\omega V_\phi+\ell)^2\right)}{r^2}\right]\phi=0.\label{eq:llmkg}
\eea
We need to specify the UV and IR boundary conditions. In the UV, i.e. near the AdS boundary at $r\to\infty$ (equivalently, $\xi=\pm\infty$), eq.~(\ref{eq:llmkg}) becomes analytically solvable, yielding:
\be
\phi\left(r\to\infty,\theta\neq 0\right)=\left(A_+r^{\Delta_+}+A_-r^{\Delta_-}\right)J_{\ell+\omega/2}\left(\frac{\pi}{2}-\theta\right),~~\Delta_\pm=\frac{1}{2}\left(1\pm\sqrt{1+4c^2}\right).\label{eq:llmkguv}
\ee
Since the $A_+$ branch diverges when $r\to\infty$ while the other branch remains finite, we put $A_+=0$ for a normalizable solution that we seek. In the IR, for $\theta=0$, it is crucial to differentiate between black-and-white and grayscale solutions. The former are smooth and the wavefunction is also to remain smooth, with a finite flux. Expanding the equation of motion for $r\to 0$, the coordinates separate again and the smooth branch is:
\be
\phi(r,\theta\to 0)=\frac{r^{\Delta_+}}{\theta}J_{\sqrt{\ell^2+2\ell-1}}(C\theta).\label{eq:kgir}
\ee
The other branch, with the Bessel function of the second kind $K$, diverges at $\theta$ small and has to be put to zero. Notice that the above expression is indeed smooth despite the $1/\theta$ term as the ratio $J_{\sqrt{\ell^2+2\ell-1}}(C\theta)/\theta$ is finite (in fact vanishing) for any integer $\ell$ for $\theta$ small. The only free parameter $C$ is obtained by matching to the UV expansion.

The grayscale case has a curvature singularity which produces an infinite potential well in the effective potential. It is a nontrivial physical problem how the solution should behave there: in fact, one has to decide what the singularity means. In accordance with the ``incipient black hole`` logic, we have decided to understand it as a horizon of vanishing size, thus adopting the infalling conditions. This yields
\bea
&&\phi(r,\theta\to 0)=\frac{1}{\theta}\left(a_1J_{\delta/2}(C\theta)+a_2Y_{\delta/2}(C\theta)\right)\nonumber\\
&&\delta=\sqrt{\left(g^2-1\right)\omega^2-2(1-g)k_1^2-2(1+g)k_2^2+4(g-1)k_1-4(g+1)k_2+4},\label{eq:kgirgray}
\eea
where the infalling solution $\sim\exp(-\imath\delta\theta/2)$ is obtained for
\be
\frac{a_2}{a_1}=\frac{\pi\sec\left(\frac{\delta\pi}{2}\right)}{\Gamma \left(-\frac{1}{2}\delta\right)\Gamma\left(1+\frac{1}{2}\delta\right)}.
\label{eq:kgirinfall}
\ee
Now that we have the boundary conditions, we can solve the equation numerically. The outcome is a finite-norm wave in the bulk, which determines some state in the CFT.\footnote{At this point it is useful to give basic information on the numerics. We use the Galerkin pseudospectral method, where the $(r,\theta)$ manifold is discretized in real space, in terms of functions from some basis. The differential operators and the system of equations thus become matrices acting on unknown functions, the latter represented by real-space vectors. The basis choice tends to be important for a successful outcome: our default choice is the Gauss-Chebyshev basis over $r$ and the Fourier basis over $\theta$ but for highly complex metrics (LLM and some superstrata) we have used the basis of spherical Bessel functions over $r$ and Legendre polynomials over $\theta$ to improve precision. Since the wave equation is always linear, no iterations are necessary.}

\begin{figure}[h!]
\centering
(A)\includegraphics[width=.4\linewidth]{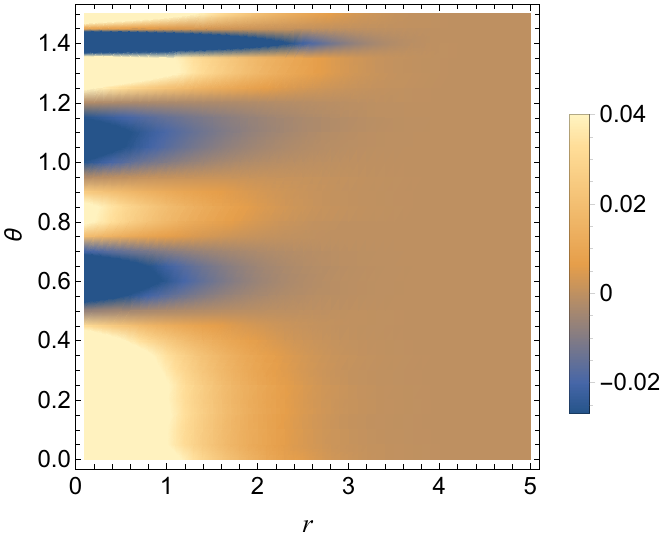}
(B)\includegraphics[width=.4\linewidth]{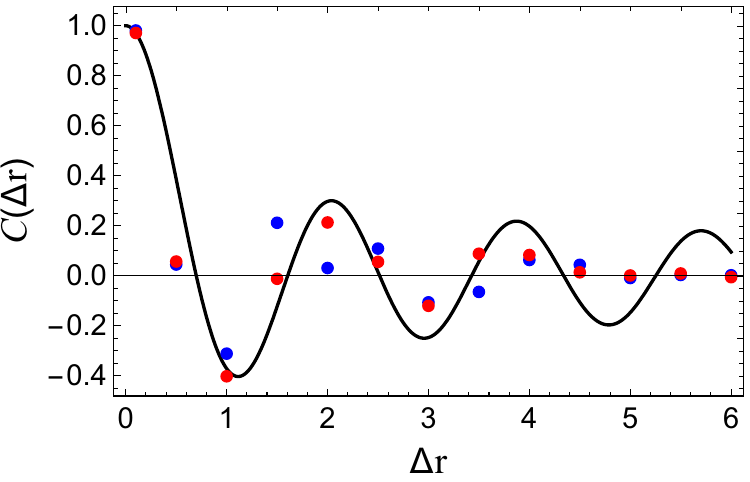}
\caption{(A) Map of wave intensity $\vert\psi(r,\theta)\vert^2$ for an exemplary solution, with $k=2$, $\tilde{k}=10$, $\ell=1$, in black-and-white disk+ring background. (B) Two-point correlation functions $C(\Delta r)$ for two backgrounds: black-and-white from panel (A), and a grayscale background. Blue points stand for the black-and-white background whereas red points stand for the gray background. Clearly, chaos is weak and the discrepancies from the Berry random wave predictions (eq.~(\ref{eq:berryconbessel}) and eq.~(\ref{eq:porterthomas}), black solid line) are systematic, but they are larger for the black-and-white, smooth solution.}
\label{fig:llmpics}
\end{figure}

A typical solution for the black-and-white background is shown in Fig.~\ref{fig:llmpics}(A). Expectedly, visual inspection of the solutions does not yield much, we give it merely for a taste of how the solutions look. A direct test of chaos is the plot of the two-point function $C(\Delta r)$, given in Fig.~\ref{fig:llmpics}(B).\footnote{For any realistic system, the correlator will depend on both points separately, i.e. we should in fact write $C(\mathbf{x}_1,\mathbf{x}_2)$, as we did in the definition (\ref{eq:corrfun}). Nevertheless, the dependence on individual points is in practice weak even when the overall agreement with the Berry form is not particularly good. Hence in this calculation we just fix one point and plot the correlator as a function of the difference.} Comparing the correlation function for both cases (black-and-white and grayscale) with the prediction of the basic Berry random wave portfolio (eq.~(\ref{eq:berryconbessel}), black curves), it becomes obvious that chaos is weak. The fit is never very good. Crucially, however, the red points, corresponding to the grayscale background, do come somewhat closer to the random wave prediction. In other words, \emph{creating a singularity or an ``incipient black hole`` makes the probe more chaotic.}

What is the reason behind the large and systematic deviations from the Bessel form of the correlator (\ref{eq:berryconbessel})? Generally speaking, they occur when the phase space (in the semiclassical limit) has a large measure of regular orbits, i.e. many KAM tori. Alternatively, it can also happen if the model is overly crude, e.g. not taking into account the strongly non-constant nature of the potential. To exclude this possibility, we have tried using the exact, position-dependent effective potential $V_\mathrm{eff}(r,\theta)$ instead of the mean value $\bar{V}_\mathrm{eff}(r,\theta)$, but the outcome is the same. Our findings thus seem robust. 


\subsection{The geodesic limit}

\begin{figure}[ht]
\centering
(A)\includegraphics[width=.4\linewidth]{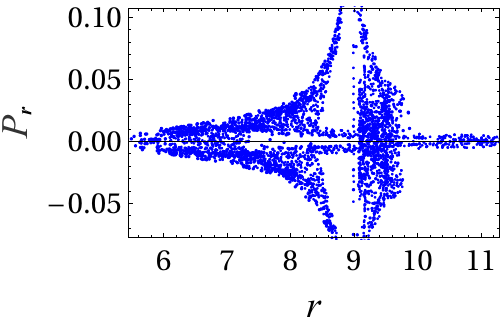}
(B)\includegraphics[width=.4\linewidth]{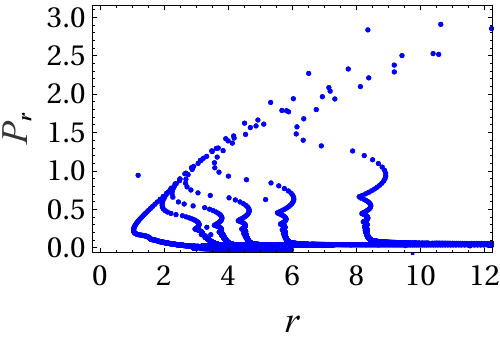}
\caption{Poincare surfaces of section for a black-and-white disk+rings background (A) and for a grayscale disk+gray ring background (B). While (A) shows typical mixed phase space with both stable structures and chaotic sea, (B) only shows narrow chaotic layers around invariant curves. The presence of the singularity almost completely destroys chaos even though the system remains nonintegrable.}
\label{fig:llmps}
\end{figure}

Having found a highly nonuniversal and very weak chaos for the wave probe, we can now look at the lightlike geodesics, which can be regarded as the high-frequency limit, i.e. the eikonal limit of the massless wave equation (\ref{eq:llmkg}). The resulting equation of motion is best written in terms of a Hamiltonian (the formal potential $V_\phi$ in given in eq.~(\ref{eq:llmmetricsph})):
\be
\mathcal H =  P_r^2 + \frac{P_{\theta}^2}{r^2} + \frac{(P_{\phi}+E V_{\phi})^2}{r^2 \cos^2 \theta} - h^4 \left( E^2 - \frac{2J_-^2}{1 - 2z} -\frac{2J_+^2}{1 + 2z} \right).\label{eq:llmHas}
\ee
This system was studied in detail in our previous work \cite{Berenstein:2025ese}, and also in \cite{Berenstein:2023vtd} in the billiard (planar) limit. We will thus just briefly recapitulate the findings from the aforementioned papers. The phase space is a typical mixed phase space, with KAM tori and stability islands surrounded by the chaotic sea. Nevertheless, at least inside the chaotic sea, the orbits are highly irregular and the Poincare section (Fig.~\ref{fig:llmps}(A)) is highly complex. 

We can thus conclude that, although the wave and geodesic dynamics are both mixed, chaos dominates in the geodesic motion while regularity dominates in the wave motion. But the most important lesson is the comparison of Fig.~\ref{fig:llmps}(A), for a black-and-white background, with Fig.~\ref{fig:llmps}(B), for a grayscale background. Here it is clear that \emph{the grayscale orbits are less chaotic and their phase space less complex}, as opposed to the situation for waves, which is exactly the other way round: waves in grayscale backgrounds are more chaotic. 

This finding -- the more chaotic the waves, the less chaotic the geodesics -- will appear repeatedly for other systems too.

\section{Mixed dynamics of 1/4-BPS and 1/8-BPS solutions in D1-D5 theory}

Now we look at the configurations with lower supersymmetry: 1/4-BPS supertubes and 1/8-BPS bubbling supertubes and superstrata. All these solutions arise within D1-D5 theory and have been proposed over the years as ``fuzzballs``, i.e. smooth, horizonless and singularity-free solutions that approximate a black hole and resolve the black hole information paradox \cite{Warner:2019jll,Bena:2025pcy}. While it would be more systematic to stay within the framework of $\mathcal{N}=4$ super-Yang-Mills theory that we studied at 1/2-BPS, we find that 1/4-BPS and 1/8-BPS solutions of super-Yang-Mills tend to be even more complex than D1-D5 fuzzballs. In addition, D1-D5 fuzzballs are very well studied and we find it more fruitful to put our results on bulk chaos in the context of systems whose behavior is in other respects known in detail.

Just as the BPS chaos papers \cite{Chen:2024oqv,Chen:2026vml} are in a sense an attempt to test the fuzzball proposal from the viewpoint of CFT state dynamics, our results can also be understood as testing the expected trend of increasing chaos for increasingly black-holish systems. In that respect, 1/4-BPS supertubes are regarded as the least black-holish (indeed, 1/4-BPS black holes are themselves ``small`` black holes with vanishing horizon area), 1/8-BPS supertubes should come closer as they have 3 charges, just like rotating D1-D5-P black holes, and 1/8-BPS superstrata are considered the best candidates as they not only have the correct number of supercharges but also more typical black-hole values of charges and momenta, and a large macroscopic entropy (but still parametrically smaller than a black hole \cite{Shigemori:2019orj}). We will see, however, that ``black-holishness`` is not so easy to define when it comes to chaos.

\subsection{1/4-BPS: supertubes}

In this paper when referring merely to ``supertubes`` we mean only the two-charge, 1/4-BPS solutions (their closest 1/8-BPS, 3-charge analogues will be called ``3-charge supertubes`` or ``bubbling supertubes``). The 2-charge supertubes were the first smooth black hole approximations to be constructed, in the form of famous Lunin-Maldacena (LM) solutions, in \cite{Lunin:2001fv,Lunin:2002iz} and other early papers; a systematic treatment from a more modern viewpoint can be found in \cite{Giusto:2019qig}.\footnote{Strictly speaking, supertubes were first constructed in the weak-coupling limit, in terms of a brane action with worldvolume fluxes; the LM solutions are the strong-coupling description of the same states, backreacting on geometry. Since we always work in the supergravity regime, we will not make this distinction.}

Since details can be found in many review papers, we will just briefly state the supertube solutions. They are two-charge states of the D1-D5 CFT, describing the degrees of freedom carried by open strings starting and ending on D1 and/or D5 branes. Being described by open strings, it is a supersymmetric gauge theory, dual to supergravity at strong coupling but well-tractable as a CFT only near the free limit (the orbifold point). In this limit, supertubes can be obtained by starting from the NS vacuum, acting on it by towers of chiral primaries, and performing the spectral flow. The result is a heavy object with a gap that decreases at high energies, bringing it closer and closer to black holes.

The supergravity solution is of the form $M_6\times\mathbb{T}^4$, the second factor being a 4-torus.\footnote{More general manifolds are found in the literature but we do not consider them.} The six-dimensional manifold $M_6$ can be written as $M_5\times\mathbb{S}^1$, with 5D supergravity determining $M_5$. For a supertube, the metric is sourced by a single winding string of arbitrary shape. As a simple but general enough example, we take the string profile to be:
\bea
&&F_1=a\cos kv,~F_2=a'\sin kv,~F_3=F_4=0\nonumber\\
&&\mathcal{F}_1=b\cos kbv,~\mathcal{F}_2=b\sin kbv,~\mathcal{F}_3=\mathcal{F}_4=0,\label{eq:stubestring}
\eea
where $\mathbf{F}$ and $\mathbf{\mathcal{F}}$ are each a four-vector, describing the string solution on $M_6$ and $\mathbb{T}$ respectively (on $M_6$ the string profile is trivial along time and along the circle $\mathbb{S}^1$). The case with $a=a'$ is the simplest, circular supertube case constructed in \cite{Lunin:2002iz}, but this case yields separable wave equation -- it is an accident stemming from the $U(1)\times U(1)$ symmetry. Any deviation from the circular symmetry of the string, i.e. any case with $a\neq a'$, brings us to the generic, non-integrable equation. We will work with $a'=2a$.

The metric on $M_6$ is then given by
\bea
ds^2=&&\frac{1}{\sqrt{\tilde{f}_1f_5}}\left[-\left(dt-\frac{a^2R}{r^2+a^2\cos^2\theta}\sin^2\theta d\varphi\right)^2+\left(dy+\frac{a^2R}{r^2+4a^2\cos^2\theta}\cos^2\theta d\psi\right)^2\right]\nonumber\\
+&&\sqrt{\tilde{f}_1f_5}\left[\left(r^2+2a^2\cos^2\theta\right)\left(\frac{dr^2}{a^2+r^2}+d\theta^2\right)+r^2\cos^2\theta d\psi^2+\left(r^2+a^2\right)\sin^2\theta d\varphi^2\right]\nonumber\\
+&&\sqrt{\frac{\tilde{f}_1}{f_5}}d\mathbf{z}^2
\label{eq:metricstube}
\eea
Here, $\mathbf{z}$ are the coordinates on the torus, $y$ is the coordinate on the circle, and $M_5$ is parametrized by the time $t$, radial coordinate $r$ and the angles $\theta$, $\psi$, $\varphi$. The redshift functions are given by:
\bea
&&\tilde{f}_1=1+\frac{Q_5k^2}{r^2+a^2\cos^2\theta}\left(a^2+b^2m^2-\frac{Q_5b^2m^2\left(\frac{a^2\sin^2\theta}{r^2+a^2}\right)^m}{Q_5+r^2+a^2\cos^2\theta}\right),\nonumber \\ 
&&f_5=\frac{a^2+2(Q_5+r^2)+a^2\cos 2\theta}{2\left(r^2+a^2\cos^2\theta\right)}.\label{eq:fstube}
\eea
The D5 brane charge is $Q_5$, the D1 brane charge is $Q_1=Qk^2(a^2+b^2m^2)$. We thus have 5 free parameters: $Q_5$, $a$, $b$, $m$ and $k$. We will fix $k=m=1$ and effectively work with $Q_5$, $a$ and $b$.

The geometry (\ref{eq:metricstube}) has three regions:
\begin{enumerate}
\item Far away (when $r\gg\sqrt{Q_1Q_5}$), the metric is asymptotically flat.
\item In the intermediate region $a^2+b^2\ll r\ll\sqrt{Q_1Q_5}$, the metric approximates AdS${}_3\times\mathbb{S}^3$. This region is holographically dual to D1-D5 CFT.
\item Finally, when $r$ becomes comparable to $a^2+b^2$, the AdS space ends with a ``cap`` as usually called -- unlike an extremal black hole where the AdS throat is infinite.
\end{enumerate}
We will focus on the probes that are mainly supported in the regions (2) and (3), as these can be related to the CFT, and to our general story of bulk vs. CFT chaos in holographic gravity. We can introduce the throat length parameter as
\be
\frak{l}\equiv\frac{\sqrt{Q_1Q_5}}{a^2+b^2}.\label{eq:throatl}
\ee
Small $\frak{l}$ corresponds to a short throat, while large $\frak{l}$ corresponds to a long throat, i.e. the black-hole-like geometry; $\frak{l}\to\infty$ yields the extremal D1-D5 black hole.

\subsubsection{Dynamics of waves and geodesics}

Now we proceed the same way as before, studying the solutions of a massless Klein-Gordon equation on the background (\ref{eq:metricstube}-\ref{eq:fstube}). Since the metric is static and independent of $\psi$, $\varphi$ and $y$, we can impose the ansatz:
\be
\Phi=e^{-\imath\omega t+\imath\ell\varphi+\imath\tilde{\ell}\psi+\imath py}\phi(r,\theta),\label{eq:stubeansatz}
\ee
yielding the partial differential equation for $\phi(r,\theta)$ that we do not state here as it is impractically long, and follows straightforwardly from the metric and the textbook definition of the Klein-Gordon equation in curved space.\footnote{Mathematica files with the equations of motion and results are available upon request.}

As we already said, D1-D5 fuzzballs become asymptotically flat at large $r$. We are not interested in this region, as it does not have a holographic dual, and dynamically it is also likely boring as it corresponds to a region of nearly flat effective potential and free wave motion. In order to zoom in into the AdS region, we expand the metric (\ref{eq:metricstube}) in $\frak{k}$ small, obtaining the decoupling limit with AdS asymptotics. All our calculations, for both waves and geodesics, are done in this metric. The boundary conditions are analogous to the black-and-white LLM case. The asymptotic behavior in the UV is:
\be
\phi(r\to\infty,\theta)=\left(A_++\frac{A_-}{r^2}+\ldots\right)\left(C_0+C_1\log\vert\cot\theta\vert+\ldots\right),\label{eq:stubeuv}
\ee
requiring the choice $A_+=0$ (no source) and $C_1=0$ (to avoid cusps and divergences). The remaining nonzero constants are unimportant as they can be absorbed in the overall normalization.
Smoothness in the IR means requiring the solution and its derivatives to be finite at the cap. This is in fact true for a generic solution, and we just need to fix the overall constant by normalization:
\be
\phi(r\to 0,\theta)\propto r^{(\ell+\tilde{\ell}-p)^2}\label{eq:stubeir}
\ee
The numerical procedure is then the same as for waves in LLM geometries.

\begin{figure}[ht]
\centering
(A)\includegraphics[width=.4\linewidth]{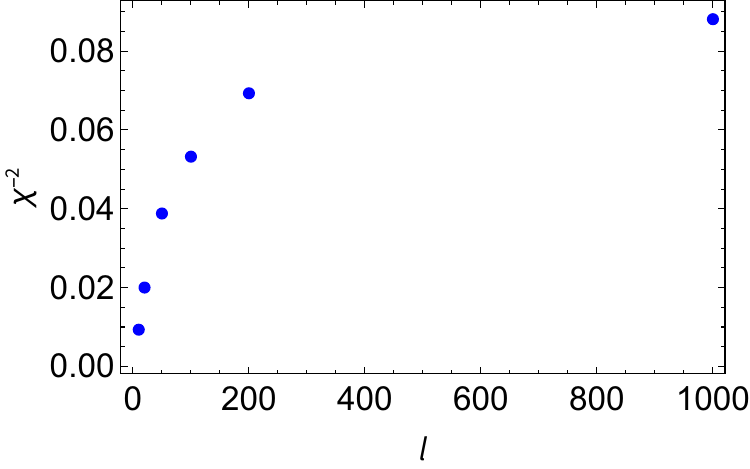}
(B)\includegraphics[width=.4\linewidth]{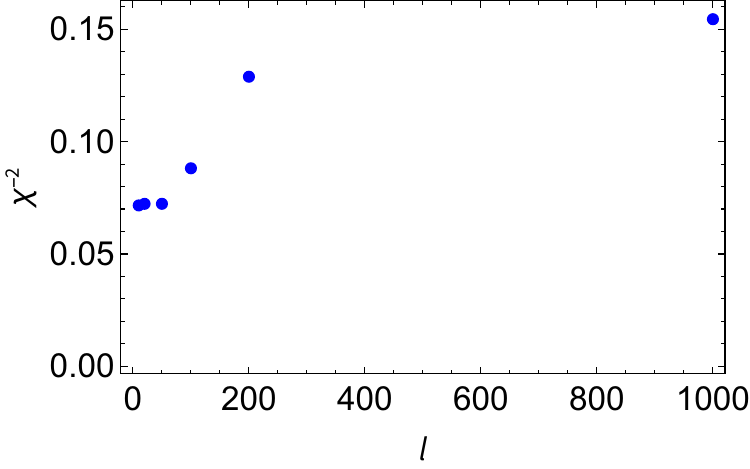}
\caption{(A) Relative proximity of the two-point function $C(\Delta\mathbf{r})$ at $\Delta\theta=0$ to the Berry random wave prediction (\ref{eq:berryconbessel}) for increasing throat lengths $\frak{l}=\sqrt{Q_1Q_5}/(a^2+b^2)$. (B) The same test for the distribution of intensities with respect to the Porter-Thomas distribution (\ref{eq:porterthomas}). In both cases, the chaos gets stronger as the throat becomes longer and longer.}
\label{fig:tpstubes}
\end{figure}

Let us first look at the two-point correlation function. As we see in Fig.~\ref{fig:tpstubes}(A), the Bessel form (\ref{eq:berryconbessel}) only comes close to the numerically computed correlators when the throat length $\frak{l}$ is long enough. For small $\frak{l}$ the correlators are far from the random wave regime. The same conclusion is reached by comparing the intensity distribution $P(I)$ to the Porter-Thomas distribution (Fig.~\ref{fig:tpstubes}(B)). The proximity of the two-point function to the form predicted by Berry is quantified by computing the inverse sum of squared deviations (normalized by the number of points $M$ and the computational unit of energy $\frak{a}$):
\be
\chi^{-2}\equiv\frac{M-1}{\frak{a}^2}\frac{1}{\sum_{j=1}^M\left[C\left(\Delta r_j\right)-C_\mathrm{BRW}\left(\Delta r_j\right)\right]^2}.\label{eq:chisquare}
\ee
This is \emph{not} the textbook Pearson chi-square test, because the null hypothesis is not a multinomial distribution but a specific model. We use the designation $\chi^{-2}$ as it is still based on normalized (inverse) sum of squared deviations but it is not a formal statistical test. Still, it clearly shows the increasing wave chaos of increasingly black-holish solutions. For $\frak{l}\geq 100$, the chaos is stronger (according to this criterion) then in any LLM geometry. Similar holds for the intensity distribution as compared to the Porter-Thomas form. This fits with the expectation that progressive decrease of supersymmetry increases chaos. 

Does the geodesic dynamics go hand-in-hand with this finding, or not (similar to the dychotomy strong geodesic chaos -- weak wave chaos found for the LLM case)? Writing the lightlike geodesic Hamiltonian akin to (\ref{eq:llmHas}), that we again leave out for brevity, we find Poincare surfaces of section with much weaker chaos than for the case of LLM. What is more, \emph{the area of the chaotic sea in the Poincare sections decreases with $\frak{l}$, unlike the wave chaos which becomes stronger.} We thus have another clear case of the counterintuitive behavior of geodesics. 


At first glance, the opposite trends in geodesics and waves go against the logic of quantum-classical correspondence: quantum chaos is precisely about the systems that are chaotic in the classical limit. On a more rigorous level, this follows from summing over periodic orbit contributions to obtain the quantum action. We propose that the reason for the unusual behavior in our systems is that, although the measure of the chaotic sea increases with increasing $\frak{l}$ and/or reducing supersymmetry, the set of stable periodic orbits at the same time becomes larger and larger (but still of measure zero, as it is discrete). The presence of such orbits is known \cite{Berry:1977,Stockmann:book} to introduce caustics in the wavefunction, which are highly localized and may coexist even with strong wave chaos -- they appear as strong but \emph{local} deviations from the random wave picture.

\begin{figure}[!ht]
\centering
(A)\includegraphics[width=.45\linewidth]{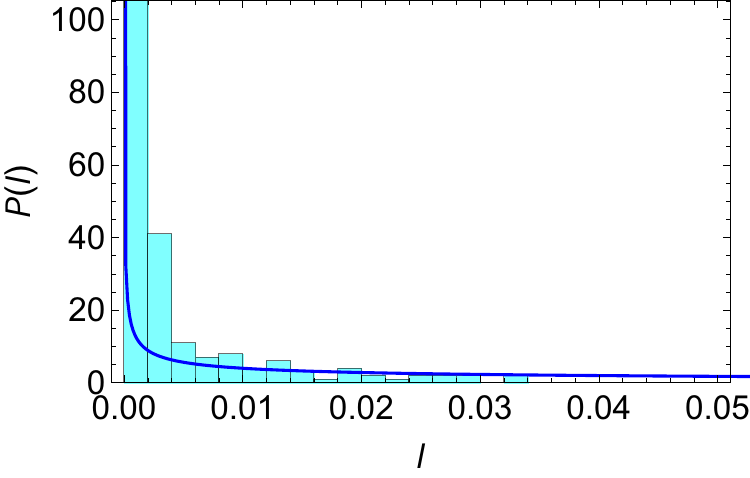}
(B)\includegraphics[width=.45\linewidth]{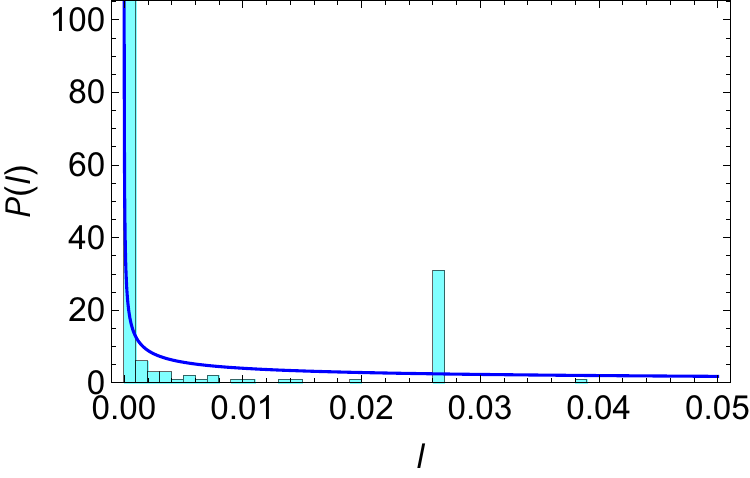}
(C)\includegraphics[width=.45\linewidth]{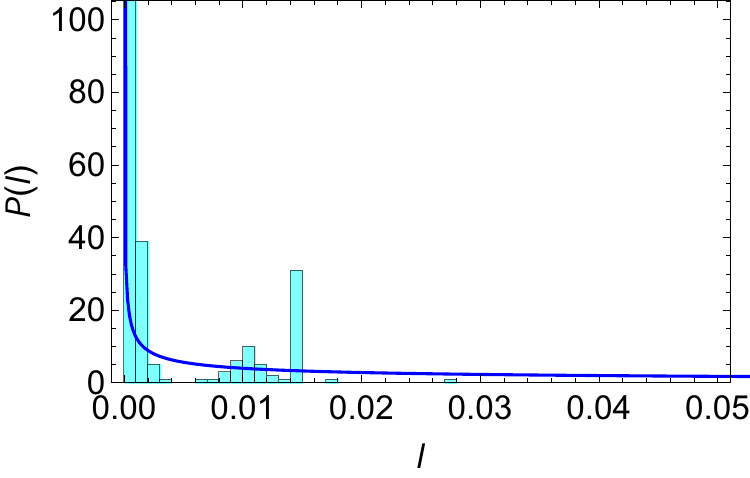}
(D)\includegraphics[width=.45\linewidth]{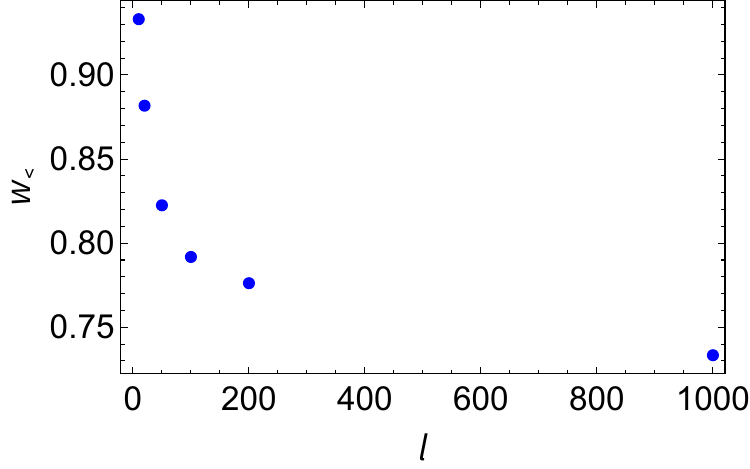}
\caption{Histograms of the intensity distribution $P(I)$ together with the prediction for strong uniform chaos, the Porter-Thomas distribution (eq.~\ref{eq:porterthomas}, magenta curves), for waves in the 2-charge supertube geometry. The throat length is $\frak{l}=10,100,1000$ (A-C). While the short-throat case (A) exhibits no significant deviations from the Porter-Thomas prediction, the cases (B-C), with $\frak{l}\gg 1$, exhibit typical signs of caustics. In the panel (D) we plot $w_<$, defined in eq.~(\ref{eq:tpweight}), for a broad range of throat lengths. The relative weight at small intensities drops with $\frak{l}$, thus caustics become more and more prominent, despite the fact that the whole solution is more and more chaotic.}
\label{fig:tpstubeshist}
\end{figure}

In Fig.~\ref{fig:tpstubeshist}(A-C) we directly observe the presence of high-$I$ spikes. Notice how these are more and more prominent as the throat grows longer: although the \emph{total mean-squared distance} from the Porter-Thomas prediction is smaller and smaller (because the low-$I$ behavior is in increasingly good agreement with the random wave prediction), the few deviations at large $I$ become more and more significant. In panel (D) we give a quantitative measure for the presence of caustics. Since caustics are rare but high spikes at large $I$, i.e. loci of high intensity, their presence can be quantified by choosing a small (but in principle arbitrary) cutoff intensity $I_0$, and computing the total weight of the intensity below $I_0$:
\be
w_<(I_0)=\int_0^{I_0}dIP(I).\label{eq:tpweight}
\ee
Clearly, small $w_<$ implies a large number and/or intensity of caustics. Large $w_<$ implies few deviations from the Thomas-Porter prediction and few caustics.

In Fig.~\ref{fig:tpstubeshist}(D) we plot $w_<$ for a sequence of throat depths, clearly finding how the presence of caustics increases with increasing throat length. We thus find a seemingly perverse outcome: while the wave chaos becomes stronger as we approach the black hole geometry (in rough agreement with the expectations of the ``black hole implies chaos`` logic), the geodesic chaos actually decreases at the same time. Let us pause and think what this could mean.

The difference between the two chaos indicators, the two-point correlation function $C(\Delta\mathbf{r})$ and the intensity distribution $P(I)$, stems from the fact that the former is nonlocal in real space: it describes correlations between pairs of points. As such, it cannot say much about orbits, which tend to be roughly localized at some mean radius $\bar{r}$: instead, it detects the presence of KAM tori. Since a KAM torus is localized in \emph{action} space, it is typically not localized around any fixed $\bar{r}$ in \emph{real} space (this is also stated in the original work \cite{Berry:1977}). On the other hand, the intensity $I=\vert\psi(\mathbf{r})\vert^2$ is defined at a single point and thus provides a strictly local measure of chaos, that does not directly see KAM tori but sees individual stable periodic orbits which act as caustics in the semiclassical quantization. In conclusion, while both the correlation function (\ref{eq:berryconbessel}) and the Porter-Thomas distribution (\ref{eq:porterthomas}) arise as consequences of the Berry random wave hypothesis, their violations signify different things: the existence of (not very complicated\footnote{See the footnote \ref{footkam}.}) KAM tori, or the existence of stable periodic orbts.\footnote{The latter phenomenon is known to be the root of quantum-mechanical or single-body scars. This is a potentially interesting topic, which is however beyond the scope of the current work.}

\subsection{1/8-BPS: bubbling supertubes and superstrata}

\subsubsection{Supergravity solutions}

In the class of 1/8-BPS systems we will consider two configurations: bubbling supertubes (3-charge supertubes) and superstrata. The 3-charge supertubes are a minimal 3-charge generalization of 2-charge ones: the torus remains untouched but the AdS throat is further deformed to produce the momentum on the circle. This breaks half of the remaining supersymmetries and gives a 1/8-BPS system. These are multi-center solutions in supergravity and can be understood as adding ``bubbles`` supported by gauge field flux to the original supertube. Out of the many constructions in the literature \cite{Bena:2004jw,Giusto:2011fy,Giusto:2012yz,Giusto:2015dfa,Giusto:2019qig} we have opted for \cite{Giusto:2012yz} which deforms the cap, i.e. the far IR of the geometry, the region most relevant for probes of chaos (in particular the geodesics).

The CFT construction outlined in \cite{Giusto:2012yz} is performed through a fractional spectral flow, which allows for a broad class of deformations that produce a momentum charge. Inspired by this logic, \cite{Giusto:2012yz} constructs the smooth supergravity solution with 3 charges that is related (if not exactly dual) to the CFT state obtained by the fractional spectral flow deformation. The metric reads:
\bea
ds^2=&&\frac{1}{h}\left(-dt^2+dy^2\right)+\frac{Q_P}{hf}\left(dt-dy\right)^2+hf\left(\frac{dr^2}{r^2+a^2(\gamma_1+\gamma_2)^2\eta}+d\theta^2\right)\nonumber\\
&&+h\cos^2\theta\left[r^2+a^2\gamma_1(\gamma_1+\gamma_2)\eta-\frac{Q_1Q_5a^2\left(\gamma_1^2-\gamma_2^2\right)\eta\cos^2\theta}{h^2f^2}\right]d\psi^2\nonumber\\
&&+h\sin^2\theta\left[r^2+a^2\gamma_2(\gamma_1+\gamma_2)\eta+\frac{Q_1Q_5a^2\left(\gamma_1^2-\gamma_2^2\right)\eta\sin^2\theta}{h^2f^2}\right]d\varphi^2\nonumber\\
&&-\frac{2\sqrt{Q_1Q_5}a}{hf}\left(\gamma_1\cos^2\theta d\psi+\gamma_2\sin^2\theta d\varphi\right)(dt-dy)\nonumber\\
&&-\frac{2\sqrt{Q_1Q_5}(\gamma_1+\gamma_2)\eta}{hf}\left(\cos^2\theta d\psi+\sin^2\theta d\varphi\right)dy\nonumber\\
&&+\frac{Q_Pa^2(\gamma_1+\gamma_2)^2\eta^2}{hf}\left(\cos^2\theta d\psi+\sin^2\theta d\varphi\right)^2+\sqrt{\frac{H_1}{H_5}}d\mathbf{z}^2.\label{eq:3stubemetric}
\eea
The coordinate system is the same as for the metric of the 2-charge supertube, with the time $t$, 1-circle $y$, angles $\theta$, $\varphi$, $\psi$ and the 4-torus $\mathbf{z}$. The metric functions are given by
\bea
&&f=r^2+a^2(\gamma_1+\gamma_2)\eta\left(\gamma_1\sin^2\theta+\gamma_2\cos^2\theta\right)\nonumber\\
&&H_1=1+\frac{Q_1}{f},~~H_5=1+\frac{Q_5}{f},~~h=\sqrt{H_1H_5}\label{eq:3stubefuns}
\eea

The number of free parameters is now 5: the charges $Q_{1,5}$, the throat-length parameter $a$ and $\gamma_{1,2}$ that are related to the fractional spectral flow parameters as $\gamma_1=-s/k$, $\gamma_2=(s+1)/k$, with $s,k$ integer. The third charge $Q_P$ is fully determined by the spectral flow parameters as $Q_P=-a^2\gamma_1\gamma_2$, and $\eta$ is not independent but is a function of charges: $\eta=Q_1Q_5/(Q_1Q_5+Q_5Q_P+Q_PQ_1)$. In practice, we will fix both $Q_{1,5}$ and $\gamma_{1,2}$ and vary only $a$, which still defines the throat length analogously to eq.~(\ref{eq:throatl}):
\be
\frak{l}=\frac{\sqrt{Q_1Q_5}}{a^2}\label{eq:throatl3}
\ee
Just like other fuzzball geometries, the bubbling supertubes have three naturally defined regions: the asymptotically flat UV region, the intermediate region, and the deep IR region where the spacetime caps off. The intermediate region for the metric (\ref{eq:3stubemetric}) is a perturbed AdS${}_3\times\mathbb{S}^3$. We are again interested in the decoupling limit, when the flat region is scaled away and we work with AdS asymptotics. The decoupling limit is performed by expanding in $\frak{l}$, and the outcome is:
\bea
ds^2=&&-\frac{f}{\sqrt{Q_1Q_5}}\left(dt^2-dy^2\right)+\frac{Q_P}{\sqrt{Q_1Q_5}}\left(dt-dy\right)^2\nonumber\\
&&-2a\left(\gamma_1\cos^2\theta d\psi+\gamma_2\sin^2\theta d\varphi\right)(dt-dy)-2a(\gamma_1+\gamma_2)\left(\cos^2\theta d\psi+\sin^2\theta d\varphi\right)dy\nonumber\\
&&+\sqrt{Q_1Q_5}\left(\frac{dr^2}{r^2+a^2(\gamma_1+\gamma_2)^2}+d\theta^2+\cos^2\theta d\psi^2+\sin^2\theta d\varphi^2\right).\label{eq:stube3ads}
\eea

The second 1/8-BPS system we study is the superstratum, a large family of solutions depending on 3 quantum numbers constructed via the holographic dictionary from explicit 1/4-BPS states of the D1-D5 orbifold CFT. The geometry is built within six-dimensional supergravity ansatz (reducing type IIB on $\mathbb{T}^4$) where the fields are determined by a set of layer equations -- a hierarchy of linear PDEs which can be solved analytically. The solutions are smooth and horizonless by construction. They are the most general microstate geometries constructed within the fuzzball program so far. In this work we focus on the $(k,m,n)=(2,1,n)$ superstrata. This family was studied, among other works, in \cite{Bena:2017upb} where it was shown that geodesic equations are non-separable and thus investigating chaos makes sense. This is believed to be the generic case; the one known example of supertubes with integrable geodesics, the $(1,0,n)$ superstratum, also studied in \cite{Bena:2017upb}, is likely a very special case. 

For the superstratum we will only give the metric in the decoupling limit, as the full metric is a mess where one can hardly gain any intuition by visual inspection. The decoupling limit is somewhat more tractable. The outer region is again flat and the inner region is again a cap, but the intermediate region is a bit more complicated than before: it interpolates between a BTZ$\times\mathbb{S}^3$ and AdS${}_3\times\mathbb{S}^3$. Explicitly, it reads \cite{Bena:2017upb}:
\bea
ds^2=&&\sqrt{Q_1 Q_5}\Lambda\Bigg[\frac{dr^2}{(r^2+a^2)}+\frac{2r^2(r^2+a^2)}{a^4R_y^2}dv^2-\frac{2(a^4(du+dv)+(2a^2+b^2)r^2H_1dv)^2}{a^4(2a^2+b^2)^2R_y^2H_2}\Bigg]\nonumber\\
&&+\sqrt{Q_1 Q_5}\Bigg[\Lambda d\theta^2+\frac{H_2}{4\Lambda}\left(d\psi+\hat {A}^{(\psi)}\right)^2+\frac{H_2\cos(2\theta)}{4\Lambda}\left(d\psi+\hat{A}^{(\psi)}\right)\left(d\varphi+\hat{A}^{(\varphi)}\right)\nonumber\\
&&~~~~~~~~~~~~~~~~+\frac{\cos^2(2\theta)H_2+\sin^2(2\theta)}{4\Lambda}\left(d\varphi+\hat{A}^{(\varphi)}\right)^2\Bigg].\label{eq:sstratummetric}
\eea 
The functions in the metric are:
\bea
&&\hat{A}^{(\psi)}=\frac{2\sqrt{2}}{R_y}\left[-\frac{dv}{2}+\frac{\cos(2\theta)(1-H_2)}{H_2} \left(\frac{a^2(du+dv)}{2a^2+b^2}+\frac{r^2H_1}{a^2}dv\right)\right]\nonumber\\
&&\hat{A}^{(\varphi)}=\frac{\sqrt{2}}{R_y}\frac{(2a^2du-b^2H_0dv)}{2a^2+b^2}\nonumber\\
&&H_0=1-\frac{r^{2n+2}}{(r^2+a^2)^{n+1}}, \nonumber\\
&&H_1=1+\frac{a^2b^2}{2(2a^2+b^2)}\frac{r^{2n}}{(r^2+a^2)^{n + 1}}, \nonumber\\
&&H_2=1-\frac{a^4b^2(n+1)}{2(2a^2+b^2)}\frac{r^{2n}}{(r^2+a^2)^{n+2}}\nonumber\\
&&\Lambda=\sqrt{1-\frac{2(n+1)a^4b^2}{(2a^2+b^2)}\frac{r^{2n}}{(r^2+a^2)^{n+2}}\sin^2\theta\cos^2\theta}\label{eq:sstratumjunk}.
\eea
The throat length can be expressed as
\be
\frak{l}=\frac{\sqrt{Q_1Q_5}}{aR_y},\label{eq:sstratuml}
\ee
so small $a$ at fixed $R_y$ is a black-hole-like geometry. The number of free parameters is 5: $Q_1$, $Q_5$, $a$, $b$, $n$, as $R_y$ can be expressed in terms of the charges. But we do not attempt to explore the whole parameter space, as we will mainly be concerned how things change as a function of $\frak{l}$. Now we have fully settled the scene and can study the dynamics of probes.

\subsubsection{Dynamics of geodesics and waves}

\begin{figure}[ht]
\centering
(A)\includegraphics[width=.4\linewidth]{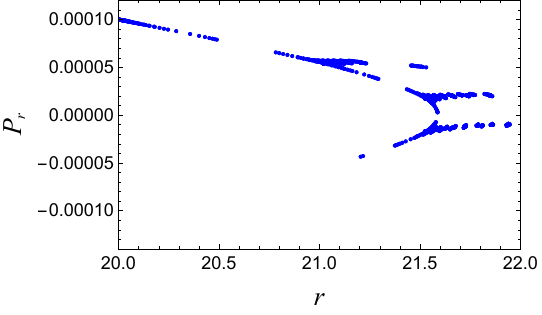}
(B)\includegraphics[width=.4\linewidth]{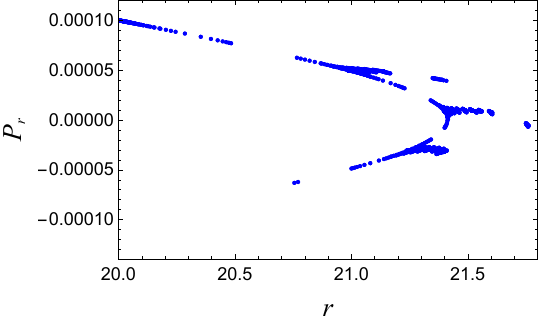}
(C)\includegraphics[width=.4\linewidth]{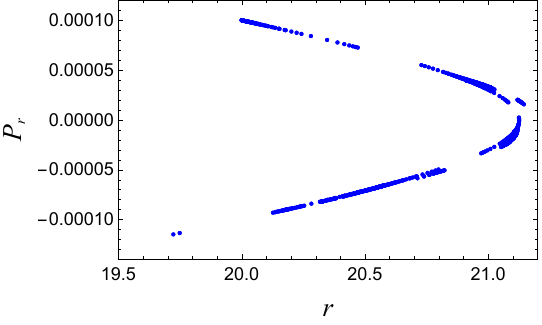}
(D)\includegraphics[width=.4\linewidth]{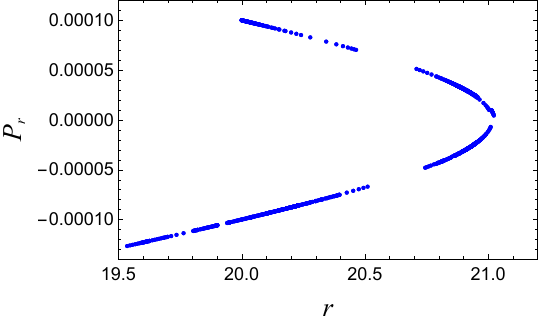}
\caption{Poincare sections for increasing throat length $\frak{l}=16,20,40,100$ (A-D). The figure (D) with the longest throat is exactly the one for which just a single KAM torus survives, with no chaotic layer.}
\label{fig:sstrataPS}
\end{figure}

As we have mentioned, the nonintegrability of geodesic motion and the non-separability of the scalar wave equation for the $(2,1,n)$ family of superstrata solutions has been established in \cite{Bena:2017upb}. Importantly, it was later noticed in \cite{Bena:2020yii} that the wave equation can become ``effectively separable`` in the limit of long AdS${}_2$ throat, meaning that for the infinitely long throat the equation would become separable -- and this is precisely the black hole limit. Running a bit ahead, we can confirm this with numerical studies of the Poincare section for null geodesics moving in this background (see Fig. \ref{fig:sstrataPS}). In this sense, the remark from \cite{Bena:2020yii} is an important hint at the physics that we find. In general, integrability of geodesics is closely related to the separability of the wave equation \cite{Frolov:2017kze} and nonintegrable geodesics imply nonseparability for waves (of course, as we have seen, it does not mean that the strength of chaos has to be comparable). For the 3-charge supertube, we have found the geodesic equation to be non-separable by direct inspection: the Hamilton-Jacobi equation does not admit the separation of variables. We are not aware of any direct nonseparability proof for the wave equation but given the geodesic nonintegrability, separability is highly unlikely.

As usual, we first sum up the boundary conditions for the scalar field $\phi(r,\theta)$ (we will again refrain from writing the several-line long wave equations). For the 3-charge supertube, the ansatz is again (\ref{eq:stubeansatz}) and $\phi(r,\theta)$ has the same leading-order behavior in both UV and IR, so conditions (\ref{eq:stubeuv}-\ref{eq:stubeir}) still apply. The superstratum gets a bit more complicated. The ansatz (already used in \cite{Bena:2020yii}) is
\be
\Phi=\exp\left[\imath\frac{\sqrt{2}}{R_y}(\omega u+pv)+\imath q_1\phi_1+\imath q_2\phi_2\right]\phi(r,\theta).
\ee
The UV behavior is again the same as in eq.~(\ref{eq:stubeuv}), as expected (the asymptotics are always AdS${}_3$). The IR solution is different:
\bea
&&\phi(r\to 0,\theta)=\cot\theta\sin^{1+q_1}\theta\cos^{1-q_2}\theta\Big[A~{}_2F_1\left(\frac{q_1-q_2}{2},1+\frac{q_1-q_2}{2},1-q_2,\cos^2\theta\right)\nonumber\\
&&+B~{}_2F_1\left(-\cos^2\theta\right)^{q_2}\left(\frac{q_1+q_2}{2},1+\frac{q_1+q_2}{2},1+q_2,\cos^2\theta\right)\Big]\label{eq:sstratumir}
\eea
The second branch has a divergent derivative, therefore we put $B=0$.

\begin{figure}[!ht]
\centering
(A)\includegraphics[width=.4\linewidth]{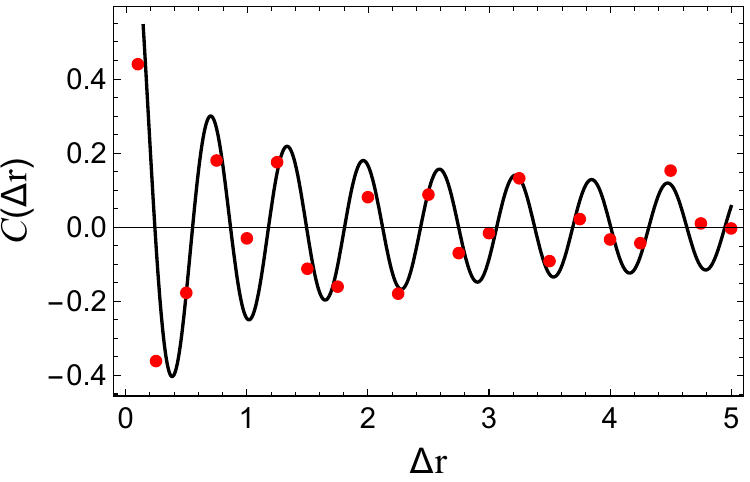}
(B)\includegraphics[width=.4\linewidth]{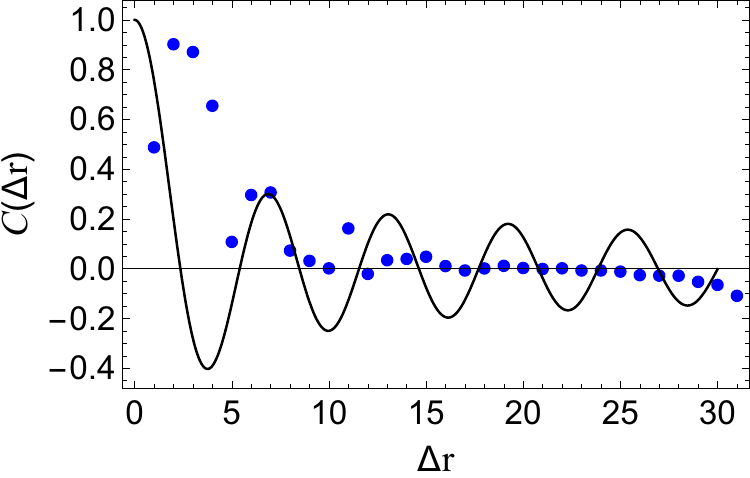}
\caption{Behavior of two-point correlation functions $C(0,\mathbf{r})$ at $\Delta\theta=0$ (points) compared to the Berry random wave prediction (eq.~(\ref{eq:berryconbessel}); solid line), for a supertube background with $n=100, R_y=10, Q_1=Q_5=100$ and $a=0.1$ (A) vs. $a=10$ (B). The case (A), with a long throat approximating a black hole, is clearly much closer to the Berry random wave regime.}
\label{fig:corrsstrata}
\end{figure}

We first check the wave chaos by comparing the two-point function with the random wave prediction (Fig.~\ref{fig:corrsstrata}). In the panel (A) we choose a very black-holish background, with a long throat, while in (B) we look at the opposite case, with a short throat. The former case follows the Berry random wave prediction to high accuracy, indeed the best of all examples we have considered. The latter case gives very poor agreement. So again, \emph{wave chaos goes hand-in-hand with the proximity of the metric to a black hole}. 

The study of Poincare sections for geodesic motion (Fig.~\ref{fig:sstrataPS}) reveals the opposite picture. All Poincare sections are very close to regular, almost fully foliated into invariant curves, with just a small chaotic region for \emph{short} throats. So the dichotomy between waves and geodesics is still present. 

\begin{figure}[ht]
\centering
(A)\includegraphics[width=.4\linewidth]{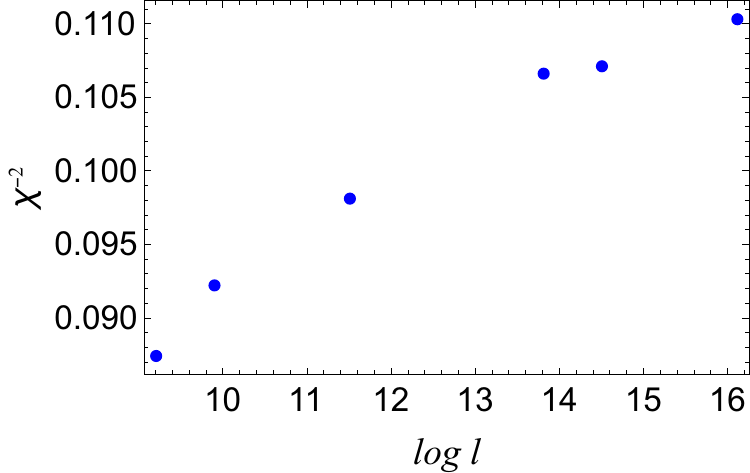}
(B)\includegraphics[width=.4\linewidth]{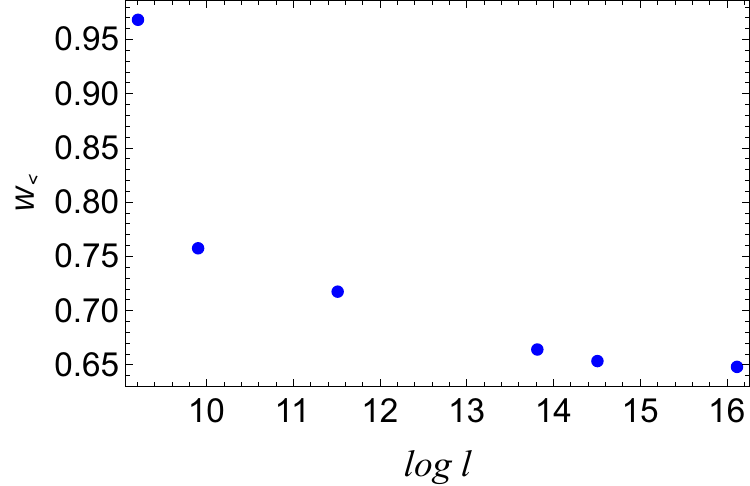}
\caption{(A) Relative inverse mean-square distance of the two-point function $C(0,\mathbf{r})$ at $\Delta\theta=0$ to the Berry random wave prediction (\ref{eq:berryconbessel}) for increasing throat lengths $l=Q_1Q_5/a^2$ of the superstratum. (B) The fraction of the spectral weight $w_<$ (defined in eq.~(\ref{eq:tpweight})) at large intensities $I>I_0$ for for increasing throat lengths $\frak{l}$ of superstrata.}
\label{fig:tpsstratum}
\end{figure}

The phenomenological explanation is again that the measure of invariant tori shrinks but the number of individual stable periodic orbits grows as the throat length grows. This is seen in Fig.~\ref{fig:tpsstratum}, showing both the proximity of the correlation function to the random wave form, in panel (A), and the percentage of spectral weight in the low-frequency domain (eq.~(\ref{eq:tpweight})), in the panel (B). We can again relate the decreasing geodesic chaos to the presence of caustics.

Similar conclusions are reached for the 3-charge supertube, except that even for long throats, the two-point correlation function is slightly less consistent with the Berry form (\ref{eq:berryconbessel}). Nevertheless, the trends are the same.

\begin{figure}[ht]
\centering
(A)\includegraphics[width=.4\linewidth]{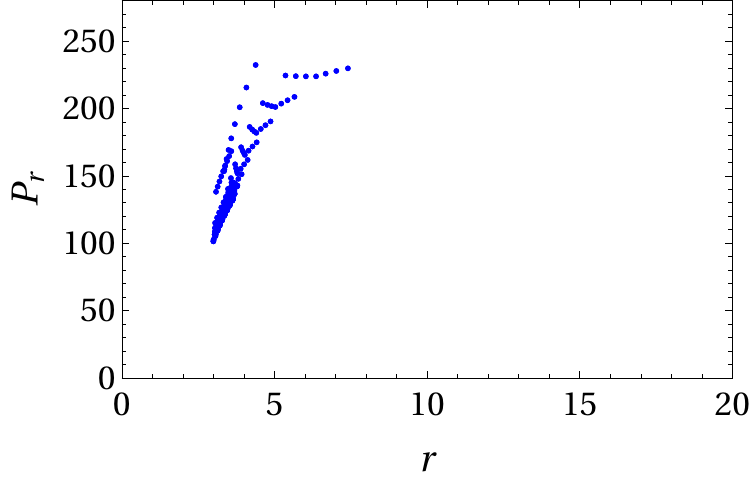}
(B)\includegraphics[width=.4\linewidth]{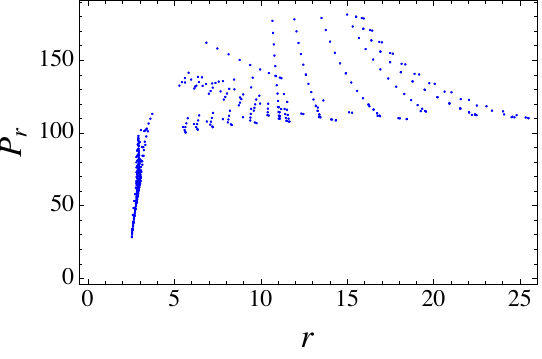}
\caption{Poincare sections for a 3-charge supertube geometry, with a long throat (A; $\frak{l}=1/\sqrt{2}$) and with a short throat (B; $\frak{l}=1/\sqrt{200}$). Similar to the superstratum solutions, there is no large chaotic sea, in contrast to relatively strong wave chaos. While the chaotic sea is absent in both cases, we can see that the geodesic in the long throat explores a smaller part of the phase space, suggesting its dynamics is more regular.}
\label{fig:ps3stube}
\end{figure}

In Fig.~\ref{fig:ps3stube} we show the Poincare sections for two throat lengths, where again most of the orbits are on invariant tori or stability islands and only very small areas are covered by chaos.

\subsection{Insights from gravitational computations}

By now we have firmly established the hierarchy of chaos. The trends are clear: approaching a black hole by reducing supersymmetry, extending the AdS throat or developing a curvature singularity all strengthen the wave chaos and weaken the geodesic chaos. We have found the phenomenological explanation for the opposite behavior of geodesics vs. waves: the reason is that geodesics are dominated by the stable periodic orbits, whose number increases even though the overall wave chaos grows stronger. 

But do we understand even the main trend? For example, why should one expect to have more chaos in the sense of the Berry random wave conjecture as the length of the AdS${}_2$ throat gets longer? We can imagine a simplified model in which a wave is captured in a cavity of depth $\frak{l}$. The spectrum of modes in the cavity will be quantized in units $\delta\omega\sim 1/\frak{l}$ and therefore the density of states will grow linearly with $\frak{l}$: $\rho\sim\delta n/\delta\omega \propto\frak{l}$. So, as the throat gets longer the total number of modes that contribute to the wavefunction grows as $N_\mathrm{throat}(\frak{l})\propto\rho\propto\frak{l}$. For $\frak{l}$ large, such a sum of many oscillatory modes will, according to the central limit theorem, behave as a Gaussian random variable. This is the intuition behind the Berry conjecture in this context.

\section{From bulk to boundary: a CFT look}

The most exciting but also the most challenging task is to relate our bulk findings to the actual microstate structure. What is currently known about the states comes from the CFT side: the usual strategy is to compute a quantity in the weakly coupled regime of the corresponding CFT and then to rely on supersymmetry protection to relate this state to the supergravity solution. For the LLM solution, a wealth of precise results has been found \cite{Berenstein:2004kk,Berenstein:2022srd,Lin:2022wdr,Kazakov:2024ald,Anempodistov:2025maj} thanks to the high degree of supersymmetry. For 1/4-BPS and 1/8-BPS sectors of the D1-D5 brane system\footnote{As we have warned in the Introduction, one should be careful not to get confused: we always count the charges with respect to type IIB string theory. Equivalently, these are the 1/2-BPS and 1/4-BPS states in D1-D5 CFT \cite{Shigemori:2020yuo}.} the problem is more difficult, but the CFT state of some supertubes and of superstrata has been constructed in detail in \cite{Giusto:2012yz,Giusto:2015dfa,Giusto:2019qig}. 

Our strategy is now to inspect the chaoticity and complexity of these states. The burning problem of such an endeavor is that a state as such is not protected: while correlation functions can be protected at least in some cases, a state will certainly change. There is no easy way out here. We take a pragmatic viewpoint: we will diagnose the degree of chaos for the super-Yang-Mills/D1-D5 states in the free theory (free Fermi gas/orbifold point) and assume that flowing toward the strongly coupled regime can only \emph{increase} chaos but should not decrease it. Therefore, if we find any signs of chaos, we believe it safe to assume that the states of the supergravity regime are also chaotic. On the other hand, if we do not find chaos, it still does not rule it out in the supergravity regime.

The Berry conjecture that we have employed in the bulk is not convenient for immediate application in the CFT, as we do not usually work in the plane wave basis in a CFT, and so do not readily know the phases. Instead, we focus at the Shannon entropy of a state (we will also consider the participation entropy but we will find that it yields similar information as the Shannon entropy) \cite{Santos:2011,Santos:2019}. If a state is defined in some (complete and minimal) basis of size $N$ as
\be
\vert\psi\rangle=\sum_{i=1}^Nc_i\vert i\rangle,\label{eq:coeffs}
\ee
then the Shannon entropy is the well-known quantity
\be
S_\mathrm{S}\equiv S_\mathrm{Renyi}(\alpha\to 1)=-\sum_{i=1}^N\vert c_i\vert^2\log\vert c_i\vert^2,
\ee
whereas the participation entropy is
\be
S_\mathrm{p}\equiv S_\mathrm{Renyi}(\alpha=2)=-\log\sum_{i=1}^N\vert c_i\vert^4.
\ee
Both are special cases of the Renyi entropy $S_\mathrm{Renyi}(\alpha)$, with $\alpha\to 1$ and $\alpha=2$ respectively. Both are known in the literature as indicators of delocalization in Hilbert space. Clearly, the minimal value $S_\mathrm{S}=S_\mathrm{p}=0$ is achieved if and only if exactly one $c_i$ is nonzero (and thus equal $1$). The maximum is reached for a vector uniformly distributed among the basis vectors when we have $c_i^2=1/N$ for all $i$, yielding $S_\mathrm{S}\sim S_\mathrm{p}\sim\log N$. Roughly speaking, any scaling of this form is consistent with a strongly delocalized (in the Hilbert space), thermalized or equilibrated \cite{Santos:2019} state. Notice we do not use the word ``chaotic``: while thermalization/equilibration/complexity typically go hand in hand with developed chaos, they are not necessarily synonymous. We will assume that the entropies tell us something about chaos, but they do not directly probe it.

In our case, when $H=0$ due to supersymmetry, there is no time evolution of the state until perturbed (clearly, a supersymmetry-breaking probe such as a scalar wave will have time-dependent dynamics). Therefore, instead of the usual approach of looking at $S_\mathrm{S}(t)$ or $S_\mathrm{p}(t)$ and concluding about the strength of chaos from the time needed to reach saturation and the distance of the saturation value from the maximum, we will compute the constant values $S_\mathrm{S}$ and $S_\mathrm{p}$ for a given microstate and analyze its scaling with $N$. This can be intuitively understood as studying an already equilibrated state, i.e. just the saturation value. 

Unlike the bulk analysis, where we started from the most supersymmetric case and proceeded to break more and more supersymmetry, in this section we will traverse the path in the opposite direction. It is convenient to first consider the superstrata, where all the key concepts are introduced. The other cases will then be progressively simpler.

\subsection{Superstratum complexity}

The CFT state of a general $(k,m,n)$ superstratum is given in \cite{Bena:2017xbt,Giusto:2019qig}. As one could expect in the supergravity regime, it has the form of a coherent state:
\be
\psi(A_1,B_{k,m,n,0}) = \sum_{p=1}^{N/k} \left( A_1 |++ \rangle_1 \right)^{N_1} \left( B_{k,m,n,0} | k,m,n,0 \rangle \right)^p,~~~N_1+kp = N,\label{eq:stateSS}
\ee
where the second equality implements the winding constraint, the angular momentum of the strands $\vert s\rangle$ takes values from the set $s\in\lbrace ++,+-,-+,--,00\rbrace$, and the states $\vert k,m,n,0\rangle$ are states in D1-D5 CFT given explicitly in \cite{Bena:2017xbt}. Computing the norm of this state
\be 
\label{eq:normPsi}\vert\psi\vert^2=\sum_p{\vphantom{\sum}}'\frac{N!}{N_1!p!}\left[\frac{1}{k}\binom{k}{m}\binom{n+k-1}{n} \right]^p\vert A_1\vert^{2N_1}\vert B_{k,m,n,0}\vert^{2p},
\ee
where $\sum_p{\vphantom{\sum}}'$ is understood to be the sum respecting the constraint from (\ref{eq:stateSS}), and using the Stirling approximation to compute the saddle-point solution that maximizes $\vert\psi\vert^2$, one gets the average strand numbers in the coherent state:
\be \label{eq:saddleSS}
    \bar N_1 = \vert A_1 \vert^2, \quad k \bar p = \binom{k}{m}\binom{n+k-1}{n} \vert B_{k,m,n,0}\vert^2.
\ee
Expanding to second order, it is easy to show that the variance behaves as $\sigma^2=\mathcal{O}(N)$ in the large-$N$ limit. As expected from the coherent states that are dual to classical superstrata geometries, the coefficients in the strand basis obey the Gaussian distribution. Schematically, from eq.~(\ref{eq:stateSS}), taking the constraint into account, we get
\be
\label{eq:cohstateSS}\psi = \sum_p A_p |N-kp, p \rangle, \quad \vert A_p\vert^2 \sim e^{- \tfrac{(p- \bar p)^2}{2 \sigma^2}},~~~\sigma^2=\mathcal{O}(N).
\ee
One might be concerned that this distribution is too wide for a coherent state. But to judge this one should compare it with the mean $\bar{p}$. From the mapping between CFT and supergravity in \cite{Bena:2017xbt} we have that $\vert B_{k,m,n,0}\vert\sim\sqrt{N}$, so in combination with eq. (\ref{eq:saddleSS}), we conclude that $\sigma/ \bar p\sim\sqrt{N}/N\sim 1/\sqrt{N}$, so the distribution is actually sharply peaked (though of course still of finite width).

We are now ready to compute the Shannon and participation entropy. Going back to our simplified model of a wave in a cavity, we can argue that in the supergravity regime for a black-hole-like objexct, the Shannon entropy will be roughly $S_\mathrm{S}\sim\log N_\mathrm{throat}(L)$, where the $N_\mathrm{throat}(L)\propto\frak{l}$ is the number of modes that contribute to the wave function $\psi=\sum_ic_i|i\rangle$. As we have mentioned, this bound will be saturated for the case of a uniform distribution of modes $\vert\mathfrak p_n\vert^2\sim\vert c_i\vert^2/\sum_a\vert c_a\vert^2\sim 1/N_\mathrm{throat}$. The task is to estimate $N_\mathrm{throat}(\frak{l})$.

A simple estimate is obtained as follows. Dealing with a coherent state from eq.~(\ref{eq:cohstateSS}) we get:
\be
\vert\mathfrak p_p\vert^2 = \frac{A_p A_p^*}{\sum_{q} A_q A_q^*} \approx \frac{1}{\sqrt{2\pi} \sigma}e^{-\tfrac{(p-\bar p)^2}{2 \sigma^2}}.
\ee
Computing the entropy from the above expression gives\footnote{From now on, we use the symbol $\simeq$ when we want to emphasize that we drop an unimportant constant, the symbol $\sim$ when we are interested in large-$N$ scaling, and $\approx$ in other cases where the calculation is approximate only.}
\be \label{eq:SSentS}
S_\mathrm{S}=-\int_{-\infty}^{+\infty} dp~ \vert\mathfrak p_p\vert^2\log\vert\mathfrak p_p\vert^2  \simeq\frac{1}{2}\log\left(2\pi\sigma^2\right)+\mathrm{const.}
\ee
Recalling that $\sigma^2\sim\mathcal{O}(N)$ we find
\be
S_\mathrm{S}\simeq\log\sqrt{N}.\label{eq:SSentS0}
\ee
Comparing this to the maximum value $\log N$, we conclude that the effective number of modes that contribute to the wavefunction at the CFT orbifold point is $N_{\mathrm{eff}}\simeq\sqrt{N}$, so $1\ll N_\mathrm{eff}\ll N$. We regard this as typical sign of weak complexity, likely implying weak chaos: the effective number of basis states contributing to the superstratum state is large (growing with $N$ and larger than any constant) but much smaller than the total Hilbert space dimension.

We can make the above back-of-the-envelope calculation more precise by extracting the non-normalized coefficients (defined in eq.~(\ref{eq:coeffs})) $c_p$ from eq.~(\ref{eq:normPsi}):
\be
c_p^2=N!\vert A_1\vert^{2N}\frac{z^p}{(N-kp)!p!},\label{eq:SScp}
\ee
where we define
\be
z\equiv\frac{\Lambda\vert B_{k,m,n,0}\vert^2}{\vert A_1\vert^{2k}},~~\Lambda\equiv\frac{1}{k}\binom{k}{m}\binom{n+k-1}{n}.\label{eq:SSzlam}
\ee
We can now compute the norm
\be
\mathcal{N}=\sum_{p=0}^{N/k}c_p^2 = N!\vert A_1\vert^{2N}\sum_{p=0}^{N/k}\frac{z^p}{(N-kp)!p!}.\label{eq:SSn}
\ee
For the special value $k=1$ that we mainly focus on, we can further simplify the expression via the binomial theorem
\be
\sum_{p=0}^{N} \frac{z^p}{(N-p)!p!} = \frac{1}{N!} \sum_{p=0}^N \binom{N}{p}z^p = \frac{(1+z)^N}{N!}~\Rightarrow~\mathcal{N}=\vert A_1\vert^{2N}(1+z)^N.\label{eq:SSbinom}
\ee
Using (\ref{eq:SScp}-\ref{eq:SSbinom}) we can now compute the distribution of normalized coefficients $\mathfrak{p}_p$:
\be
\vert\mathfrak p_p \vert^2 = \frac{c_p^2}{\mathcal N} = \binom{N}{p} \left( \frac{z}{1+z} \right)^p \left( \frac{1}{1+z} \right)^{N-p} = \mathrm{Bin}(N,q),\label{eq:SSbin}
\ee
where $\mathrm{Bin}(N,q)$ is the binomial distribution written in the variable $q\equiv z/(1+z)$. Finally, using the known identities for the mean values of binomial distribution we get for the entropy:
\be \label{eq:SSspecA}
S_\mathrm{S}=-N\left(q\log q +(1-q)\log(1-q)\right)-\sum_p \mathrm{Bin}(N,q)\log\binom{N}{p}.
\ee
The last term can be computed in the large-$N$ limit, using the Stirling approximation:
\be\label{eq:SSspecB}
\left\langle\binom{N}{p}\right\rangle_{\mathrm{Bin}(N,q)}\approx N\left(q\log q-\left(1-q\right)\log\left(1-q\right)\right)-\frac{1}{2}\log\left(2\pi\frac{(N-p)p}{N}\right).
\ee
Evaluating this expression at the mean value of the strand number $p\mapsto\bar{p} = Nq$ we notice that the first term in eq.~(\ref{eq:SSspecA}) is canceled by the first term in eq.~(\ref{eq:SSspecB}), leaving us only with
\be
S_\mathrm{S}\simeq\frac{1}{2}\log\left(2\pi Nq\left(1-q\right)\right).\label{eq:SSentS1}
\ee
Recalling that the variance of the binomial distribution is given by $\sigma^2 = N q(1-q)$ we recover the quick-and-dirty Gaussian result from eq.~(\ref{eq:SSentS0}).

More generally, for an arbitrary value of $k$ we have to deal with the following generating functional:
\be
\mathcal Z_N(z)=\sum_{p=0}^{N/k} \frac{z^p}{(N-kp)!p!},\label{eq:SSzdist}
\ee
implying that
\bea
\mathcal N &=& \vert A_1\vert^{2N}\mathcal Z_N(z),\label{eq:SSnn}\\
\vert\mathfrak p_p\vert^2 &=&\frac{c_p^2}{\mathcal N}=\frac{1}{\mathcal Z_N(z)}\cdot\frac{z^p}{(N-kp)!p!}.\label{eq:SSpp}
\eea
The Shannon entropy is given by
\be \label{eq:SSspecEntr2}
S_\mathrm{S}=\log\mathcal Z_N(z)-\langle p\rangle\log z+\langle\log (N-kp)!\rangle+\langle\log p! \rangle,
\ee
where $\langle \cdots \rangle = \sum_p \vert\mathfrak p_p \vert^2 (\cdots)$. Using the recurrence relation $\mathcal Z_N'(z) = \mathcal Z_{N-k}(z)$ and expanding around the dominant saddle $\bar p$ in the large-$N$ behavior of $\mathcal Z_N(z)$
\be
\mathcal Z_N(z)\sim\frac{z^{\bar{p}}}{(N-kp)!p!}\cdot\sqrt{2\pi}\sigma,
\ee
where the factor $\sqrt{2\pi}\sigma$ comes from the integration measure, i.e. we promote $\sum_p \to \int_{- \infty}^{+ \infty} dp$, we can easily show that the expectation value $\langle p\rangle$ is precisely the saddle-point value $\bar p$. Expanding the first term's contribution to the entropy in this way gives
\be \label{eq:SSlogZ}
\log \mathcal Z_N(z) \approx \bar p \log z - \log(N-k \bar p)! - \log \bar p! + \frac{1}{2}\log{\left(2\pi \sigma^2\right)}.
\ee
Observing the equations (\ref{eq:SSspecEntr2}) and (\ref{eq:SSlogZ}) we again notice the leading terms cancel out, so we are only left with
\be
S_\mathrm{S}\simeq\frac{1}{2}\log\left(2\pi\sigma^2\right)=\frac{1}{2}\log\left(2\pi\sqrt{N}\right).
\ee
Equivalently, we can understand the above result through the number of states effectively participating in the microstate:
\be
S_\mathrm{S}\simeq\log N_\mathrm{eff},~~N_\mathrm{eff}=\sqrt{N}.
\ee
We might call this a weakly chaotic, or rather weakly complex state: it only spreads over $\sim\sqrt{N}$ basis vectors in a Hilbert space of dimension $N$, a fraction that grows indefinitely with $N$ but on the other hand becomes a vanishing subspace of the whole Hilbert space for $N$ large.

Instead of the Shannon entropy, we could look at the participation entropy, which is more widely used in the literature on chaos and thermalization in quantum many-body systems as a sign of state spread and thermalization. But at the level of accuracy of our computations (for large $N$ and employing the saddle-point approximation) the scaling is exactly the same. We can express $S_\mathrm{p}$ as
\be
S_\mathrm{p}=-\log\sum_p\vert\frak{p}_p\vert^4=-\langle\vert\frak{p}_p\vert^2\rangle=-\langle\mathrm{Bin}(N,q)^2\rangle,
\ee
in full analogy with eqs.~(\ref{eq:SSspecA}-\ref{eq:SSspecB}). For $k=1$,\footnote{The argument is analogous for general $k$.} making use of eq.~(\ref{eq:SSbin}) and the same identities for the binomial distribution as above, we find, at the saddle point
\be
S_\mathrm{p}=\binom{N}{\bar p}\left(\frac{z}{1+z}\right)^{\bar{p}}\left(\frac{1}{1+z}\right)^{N-\bar{p}}.
\ee
Inserting $\bar{p}=Nq$ and writing out the above expression in the Stirling approximation, we find again that the leading contributions of the form $n\log n$ cancel and only the subleading terms of the Stirling formula contribute. This finally yields
\be
S_\mathrm{p}\simeq\frac{1}{2}\log\left(2\pi Nq\left(1-q\right)\right),
\ee
the same form as for the Shannon entropy. This happens for other systems we study too, so for this reason we only consider the Shannon entropy, understanding that the (more popular) participation entropy yields compatible behavior.

\subsubsection{Implications for the structure of the correlation functions}

Finally, we can take a blitz look at the structure of the correlation functions, based on the above results. The form of the state given by eqs.~(\ref{eq:saddleSS}-\ref{eq:cohstateSS}) has profound implications for the structure of two-point correlation functions, which in the CFT language used above can be schematically represented as
\be
C_{p,q}\sim\sum_{p,q}A_p^\dagger A_q.\label{eq:corrfunss}
\ee
Bearing in mind the estimate from eq.~(\ref{eq:cohstateSS}) that $A_p$ are always of the form $\exp(-(p-\bar p)^2)$, the object (\ref{eq:corrfunss}) will clearly not show any signs of randomness. It is true that, strictly speaking, we should first perturb the coherent state to compute the correlation function that corresponds to a probe (wave or geodesic) on top of the background. The scalar wave in the bulk would correspond to a scalar operator with scaling dimension $\Delta \sim \mathcal O(1)$, such as $J_{-1}^+ J_{-1}^-$. After working out how this operator acts on the coherent state and redoing the previous exercise, one concludes that the variance will receive $\mathcal{O}(1/N^2)$ corrections on top of the previous $\mathcal{O}(N)$ result. From this we can further conclude that even if we act with $\left( J_{-1}^+J_{-1}^- \right)^N$, i.e. an operator with $\Delta \sim \mathcal{O}(N)$, the Gaussian structure will still be preserved, as the variance will receive $\sim \mathcal{O}(1/N)$ correction. Thus, no evidence for the randomness in the sense of the Berry conjecture can be found from a coherent state of the form given by eqs.~(\ref{eq:saddleSS}-\ref{eq:cohstateSS}), at vanishing 't Hooft coupling.

This need not to be in contradiction with our findings in the bulk. Based on the findings from \cite{Bena:2018bbd} one can argue that the operators that are used to build up the CFT states, namely $L_{-1}$ and $J_{-1}^3$, undergo  mixing with operators having fractional momenta, like $L_{-1/N_1N_5 }$, as we move from the orbifold point to the supergravity point in the moduli space. At the end of the day, we can reiterate what we said at the beginning of this section: not finding chaos in the weakly coupled CFT regime does not preclude it from arising at strong coupling. The entropy growing as a positive power of $N$ does yield some indication of chaos that arises at strong coupling, while the above, very rough sketch of the correlation function behavior, does not.

\subsection{Supertube complexity: 3-charge}

Let us now consider another 3-charge, 1/8-BPS state, relevant for the supertubes. To the best of our knowledge, the CFT state corresponding to the supergravity solution given in eq.~(\ref{eq:3stubemetric}) and taken over from \cite{Giusto:2012yz} has not been explicitly constructed, as the fractional spectral flow required to produce it is hard to evaluate explicitly (although \cite{Giusto:2012yz} does give a rather detailed description of what this state should look like). We will thus examine another 1/8-BPS, 3-charge state of D1-D5 CFT given in \cite{Giusto:2015dfa}, which should hopefully capture roughly the same physics\footnote{The ``microstate at the cap`` constructed in \cite{Giusto:2012yz} is built out of excitations from the bosonic sector whereas the state in \cite{Giusto:2015dfa} is built out of free fermions.} (on the other hand, this state does not have an explicitly known supergravity dual).

The microscopic state constructed in \cite{Giusto:2015dfa} is obtained by exciting different strands (component strings) $k$ by acting with the angular momentum raising operator $\left(J_{-1}^+\right)_k$:
\be
\psi\left(\{N_{k,m_k},N_k^{(s)}\}\right)=\prod_{s}\prod_{k=1}^M\left( |s\rangle_k \right)^{N_k^{(s)}} \prod_{k,m_k} \left( \frac{(J_{-1}^+)_k^{m_k}}{m_k!}|00\rangle_k \right)^{N_{k,m_k}}\label{eq:3stubestate}
\ee
The state $\vert s\rangle$ can again take any combination of angular momenta ($s\in\lbrace ++,+-,-+,--,00\rbrace$) although we could also fix it to $00$ as we did for the superstratum, without changing the final outcome. The number of strands $M$ is arbitrary and we can treat it is another free parameter of the state. The constraint for the state (\ref{eq:3stubestate}) is:
\be
\sum_{s,k}kN_k^{(s)}+\sum_kkm_kN_{k,m_k}=N.
\ee
With this constraint, the norm of the state (found from the total number of terms in the products) is:
\be
\mathcal{N}\left(\{N_{k,m_k}\},\{N_k^{(s)}\}\right)=\left( \frac{N!}{\prod_{s}\prod_{k} N_k^{(s)}! \, k^{N_k^{(s)}}} \right)\left( \frac{1}{\prod_{k,m_k} N_{k,m_k}! \, k^{N_{k,m_k}}} \right) \prod_{k,m_k} \binom{k}{m_k}^{N_{k,m_k}}.
\ee
As we have explained for the superstratum, the supergravity state again is a coherent state containing the sum of many microstates, with coefficients $A_k^{(s)}$ and $B_{k,m_k}$:
\be
\psi(\{A_k^{(s)}, B_{k,m_k}\}) = \sum_{\{N_{k,m_k}^{(s)}\}} c\left(\{N_{k,m_k}\},\{N_k^{(s)}\}\right)\left[ \prod_{s}\prod_{k}(A_k^{(s)} |s\rangle_k)^{N_k^{(s)}} \prod_{k,m_k} \left( B_{k,m_k} \frac{(J_{-1}^+)_k^{m_k}}{m_k!} |00\rangle_k \right)^{N_{k,m_k}} \right].
\ee
From the orthogonality of basis states we can express the coefficients as:
\be
c^2\left(\{N_{k,m_k}\},\{N_k^{(s)}\}\right)=\mathcal{N} \left( \prod_s\prod_{k}\left(|A_k^{(s)}|^2 \right)^{N_k^{(s)}} \right) \left( |B_{k,m_k}|^2 \right)^{N_{k,m_k}},
\ee
yielding, with the use of (\ref{eq:3stubestate}):
\be
c^2\left(\{N_{k,m_k}\},\{N_k^{(s)}\}\right)=N!\left( \prod_s\prod_{k}\frac{1}{N_k^{(s)}!} \left( \frac{|A_k^{(s)}|^2}{k} \right)^{N_k^{(s)}} \right) \left( \prod_{k,m_k} \frac{1}{N_{k,m_k}!} \left( \binom{k}{m_k} \frac{|B_{k,m_k}|^2}{k} \right)^{N_{k,m_k}} \right).
\ee
Still closely following the superstratum calculation, we define the generating function for the coefficients $\mathcal{Z}_N$ and the probability distribution function for the coefficients $P$:
\be
P=\frac{N!}{\mathcal{Z}_N} \prod_s\prod_{k}\frac{(z_k^{(s)})^{N_k^{(s)}}}{N_k^{(s)}!} \prod_{k,m_k} \frac{(z_{k,m_k}^{(0,0)})^{N_{k,m_k}}}{N_{k,m_k}!}\label{eq:3stubep},
\ee
where we have also defined:
\be
z_k^{(s)}=\frac{|A_k^{(s)}|^2}{k},\,\, z_{k,m_k}^{(00)}=\binom{k}{m_k}\frac{|B_{k,m_k}|^2}{k}.\label{eq:3stubegenz}
\ee
Now we compute the Shannon entropy as $S_\mathrm{S}=-\langle\log P\rangle$, making use of the distribution (\ref{eq:3stubep}). Using the Stirling approximation and then finding the saddle points as $\partial\log P/\partial N_k^{(s)}=$ $\partial\log P/\partial N_{k,m_k}=0$, we get:
\be
\bar{N}_{k,s}=\sigma_{k,s}^2=\frac{|A_k^{(s)}|^2}{k},~~\bar N_{k,m_k}=\sigma_{k,m_k}^2=\binom{k}{m_k}\frac{|B_{k,m_k}|^2}{k}.
\ee
In the saddle point (mean-field) approximation, the probability distribution again becomes Gaussian:
\be
P(N)\approx\prod_{s,k}\frac{1}{\sqrt{2\pi\sigma_{s,k}^2}}\exp\left(-\frac{(N_k^{(s)}-\bar{N}_k^{(s)})^2}{2\sigma_{s,k}^2} \right) \prod_{k,m_k}\frac{1}{\sqrt{2\pi\sigma_{k,m_k}^2}} \exp\left(-\frac{(N_{k,m_k}^{(0,0)}-\bar{N}_{k,m_k}^{(0,0)})^2}{2\sigma_{k,m_k}^2}\right).
\ee
Computing the Shannon entropy of the spectrum by definition, we get:
\be
S_\mathrm{S}\approx\sum_{s}\sum_{k}\left(\frac{1}{2}\log(2\pi\sigma_{s,k}^2)+\frac{1}{2}\right)+\sum_{k,m_k}\left( \frac{1}{2}\log(2\pi\sigma_{k,m_k}^2)+\frac{1}{2}\right),
\ee
where the constraints imply $N=\sum_{s}\sum_k\sigma_{s,k}^2+\sum_{k,m_k}\sigma_{k,m_k}^2$. But now a complication arises with respect to the superstratum case: the superstratum that we have considered only had two different quantum numbers which, after implementing the constraint, became only one. The state we are considering now has $M$ ``seed`` states, i.e. strands with different $k$ and the constraint does not fully determine the distribution of their occupation numbers. But it will turn out that the scaling of entropy is independent of that. Therefore, we can make a schematic assumption that the distribution over $(k,s)$ carries the proportion $\lambda$ of the total dispersion, and the distribution over $(k,m_k)$ consequently $1-\lambda$:
\be
\sum_{s}\sum_{k}\sigma_{s,k}^2=\lambda N,~~\sum_{k,m_k}\sigma_{k,m_k}^2=(1-\lambda)N.
\ee
This yields for the entropy:
\be
S_\mathrm{S}\sim\sum_{s}\sum_{k=1}^M\log\left(2\pi N\sqrt{\lambda(1-\lambda)}\right)\sim\frac{M_s(M-1)}{2}\log\left(2\pi N\sqrt{\lambda(1-\lambda)}\right).\label{eq:3stubeent}
\ee
The factor $M_s(M-1)$, that was absent in the superstratum, is in a sense fake: it comes simply from the summation over $M_s$ possible spins for the seed states (at most 5 -- $\lbrace ++,+-,-+,--,00\rbrace$) and over $M$ strands with different $k$. For the superstraum, we have fixed $s$ to $00$ and $M$ to $2$ from the beginning. To facilitate comparison with the superstratum, we therefore again fix the ``seed`` states to $s=00$ and $k$ taking two possible values, so (\ref{eq:3stubeent}) becomes
\be
S_\mathrm{S}\simeq\log\left(2\pi N\sqrt{\lambda(1-\lambda)}\right).\label{eq:3stubeent1}
\ee
which we can directly compare to the result for the superstratum.

Let us pause and think about the result (\ref{eq:3stubeent1}). The fact that we had to eliminate the prefactors of $M$ and $M_s$ by hand is trivial: remember that the superstratum states that we have looked at are very special. The Shannon entropy counts the number of directions in the Hilbert space that the state spans, therefore it has to be proportional to the size of the family of solutions we look at. The $\lambda$-dependent prefactor is also irrelevant. But the real surprise is the fact that \emph{the entropy scales as $\log N$, not $1/2\log N$ as for the superstratum}. In other words, $N_\mathrm{eff}=N$ and not $\sqrt{N}$. Therefore, this class of supertubes can reproduce the scaling expected for black holes\footnote{We again emphasize that the Shannon entropy and other Renyi entropies of states have nothing to do with the Bekenstein-Hawking entropy, which is known not to be reproduced by any currently known smooth solutions.} and in the sense of complexity is thus more black-holish than the superstratum, despite the latter having shown stronger wave chaos. This is important: the bulk and the CFT criteria (at least in this, weakly coupled regime) need not go hand in hand.

\subsection{Supertube complexity: 2-charge}

The next stop is the the 1/4-BPS supertube. In this case, we can take the proposed CFT state that corresponds to the same bulk supergravity solution that we have tested. The state and the constraint read \cite{Giusto:2015dfa}:
\be
\psi\left(\{N_k^{(s)}\}\right)=\prod_{k,s}\left(|s\rangle_k \right)^{N_k{(s)}}, \quad \sum_{k,s} k N_k{(s)}=N,
\ee
which in supergravity becomes the coherent state:
\be
\psi(\{A_k^{(s)}\})=\sum_{\{N_k^{(s)}\}}c_k^{(s)2}\prod_{k,s}\left(A_k^{(s)}|s\rangle_k\right)^{N_k^{(s)}0}.
\ee
We follow the familiar path, expressing the coefficients in terms of the generating function and writing the probability distribution for the coefficients:
\bea
c_k^{(s)2}&=&\mathcal{N}\left(\{N_k^{(s)}\}\right)\prod_{k,s}|A_k^{(s)}|^{2N_k^{(s)}}=N! \prod_{k,s}\frac{{z_k^{(s)}}^{N_k^{(s)}}}{N_k^{(s)}!}\\
\mathcal{Z}_N&=&\sum_{N_k^{(s)}}\prod_{k,s}\frac{\left(z_k^{(s)}\right)^{N_k^{(s)}}}{N_k^{(s)}!}\\
P&=&\frac{N!}{\mathcal{Z}_N} \prod_{k,s} \frac{\left( z_k^{(s)} \right)^{N_k^{(s)}}}{N_k^{(s)}!}.
\eea
From $S_\mathrm{S}=-\langle\log P\rangle$, the Stirling formula and the saddle-point approximation we get
%
\be
\bar N_k^{(s)}=\left(\sigma_k^{(s)}\right)^2=z_k^{(s)}.
\ee
Now expanding $P$ and computing the Shannon entropy by definition yields:
\be
S_\mathrm{S}\approx\sum_i\left(-\int_{-\infty}^{+\infty}P(N_i)\log P(N_i)\,dN_i \right)=\sum_{k,s}\left( \frac{1}{2}\log\left(2\pi\left(\sigma_k^{(s)}\right)^2\right)+\frac{1}{2} \right).
\ee
Similarly to the 1/8-BPS supertube state, the entropy adds up to $M_s(M-1)\times S_1$, where $S_1$ is the entropy per strand. As we have already explained, to get a meaningful comparison we fix $M_s=1$, $M=2$ to get:
\be
S_\mathrm{S}\simeq\frac{1}{2}\log\left(2\pi N\right)\sim\log\sqrt{N}\equiv\log N_\mathrm{eff}
\ee
where $N_\mathrm{eff}=\sqrt{N}$, coming back to the superstratum result.

So far, we have seen that $N_\mathrm{eff}$, i.e. the behavior of Renyi entropies, at least in the weak-coupling regime, is not as as neat and clear-cut as the chaos of supergravity probes: at 1/8-BPS we have $N_\mathrm{eff}=N$ or $N_\mathrm{eff}=\sqrt{N}$ depending on the microstate; at 1/4-BPS it is again $N_\mathrm{eff}=\sqrt{N}$, at least for this example. 

\subsection{No complexity in 1/2-BPS LLM states}

Let us finally briefly consider the 1/2-BPS states of the LLM solutions. We employ the fermionic representation of eqs.~(\ref{eq:llmfermi}-\ref{eq:fermidiskring}). Consider the disk+ring configuration. This configuration, as given by eq.~(\ref{eq:fermidiskring}), is completely fixed by the radii $R_{1,2,3}$. The state is a unique Slater determinant (at the weakly coupled level), and since the whole Hilbert space is also spanned by Slater determinants the coefficients analogous to $c_k$ or $c_k^{(s)}$ as in previous subsections equal $\delta_{i,i_0}$ where $i_0$ is the state of the given pattern. Such coefficients obviously imply zero Renyi entropy. We can also see it directly in the Z-word representation: as it is known \cite{Lin:2004nb,Anempodistov:2025maj}, the state described by the disk with any number of rings is obtained by acting on the vacuum with Schur polynomials of $Z$ operators. All the coefficients of the polynomial are fixed: localization in the radial direction (i.e., the formation of sharp black and white rings) follows automatically from the symmetry of the Schur polynomials. Therefore, there is nothing to count.

One  might worry that this is an artifact of the chosen state. Indeed, more general states, without circular symmetry, are given by coherent states of the form \cite{Berenstein:2022srd,Lin:2022wdr,Anempodistov:2025maj}:
\be
\vert\Lambda\rangle=\frac{1}{\mathrm{Vol}\,U(N)}\int dU\vert\Lambda\vert U\rangle=\frac{1}{\mathrm{Vol}\,U(N)}\int dU\exp\left[\mathrm{Tr}U\Lambda U^{-1}Z^\dagger\right]\vert 0\rangle,
\ee
where the eigenvalues of the matrix $\Lambda$ define the location of the black regions in in the LLM plane, whereas the integration over the gauge group $U(N)$ is necessary to keep the state gauge-invariant. Now the sum over coherent state coefficients becomes a continuous integral. However, there is no state-dependent Shannon entropy here, just a factor related to the volume of the gauge group -- but gauge transformations do not lead to physically distinct states. The coefficient of each microscopic state $\vert\Lambda\vert U\rangle$ is $1/\mathrm{Vol}\,U(N)$ so
\be
S_\mathrm{S}=\frac{1}{\mathrm{Vol}\,U(N)}\int dU\log\mathrm{Vol}\,U(N)=\log\mathrm{Vol}\,U(N).
\ee
We could insert the volume of the $U(N)$ group manifold in the above result and obtain a quantity behaving roughly as $N^2$. But this is merely the gauge overcounting. There is no summation over physically distinct states, and the Shannon (as well as participation) entropy is zero at 1/2-BPS.

\section{Discussion and conclusions}


At the end of the journey, we can say that we have fulfilled two out of three initial goals. We have characterized the bulk chaos in BPS backgrounds as a function of important properties (number of supercharges, proximity to black hole near-horizon geometry, smoothness) and we have found a pristine trend of increasing chaos with increasing black-holishness. Another important finding is the behavior of geodesics: their motion becomes more regular at the same time as the waves become more chaotic.

The dichotomy between the behavior of waves and geodesics is seemingly paradoxical. At a qualitative level, however, we understand the mechanism behind it. The crucial detail is that waves are nonlocal while geodesics are local, so they are sensitive to the approximate symmetries of the potential well, i.e. the black-hole-like long throat where they spend most of the time. Waves exist everywhere at the same time and are not that sensitive to local properties. The simplicity of dynamics deep inside the throat leads to wave caustics, which do not significantly change the effectively random behavior of waves, but directly lead to nearly regular geodesic motion. As we mentioned, this sounds somewhat like quantum scars in billiards and other first-quantized systems. To what extent the relation to scars can be formalized we will see in the future. We have thus answered the questions (2) and (3) from the Introduction, and the summary of trends is given in Table \ref{tab:chaos}. 

\begin{table}[ht]
\centering
\begin{tabular}{|c|c|c|c|c|c|}
\hline
& Wave $C(\mathbf{r}_1,\mathbf{r}_2)$ & Wave $P(I)$ & Caustics & Geodesic chaos\\
\hline
No. of supercharges $\uparrow$  & $\downarrow$ & $\downarrow$ & $\downarrow$ & $\uparrow$\\
AdS throat length $\uparrow$  & $\uparrow$ & $\uparrow$ & $\uparrow$ & $\downarrow$\\
Curvature singularity & $\uparrow$ & $\uparrow$ & $\uparrow$ & $\downarrow$\\
\hline
\end{tabular}
\caption{The trends in bulk chaos as a function of the fraction of supersymmetry preserved, AdS throat length and existence of the singularity. The chaos indicators considered are the chaos in the 2-point correlation function for waves, the overall shape of the intensity distribution for waves, the existence of caustics in the intensity distribution, and the strength of geodesic chaos. By $\uparrow$/$\downarrow$ we denote the increasing/decreasing trend.}
\label{tab:chaos}
\end{table}

Question 1 from the Introduction -- how to extrapolate these findings to black holes themselves, and how to find a red line between black-hole and non-black-hole backgrounds -- we have not quite answered. To the best of our knowledge, most top-down black hole solutions exhibit both integrable geodesics and separable wave equations, so at least for waves it seems the black hole limit is not smooth -- which is perhaps not surprising as horizons and singularities are nonperturbative, yes-no notions. Still, what we have done so far provides a natural stepping stone for further work. For example, we can take ensemble averages of superstrata or 3-charge supertubes and study the probe dynaics in such systems, essentially repeating the LLM story of \cite{Berenstein:2025ese} for more realistic black hole models, at lower supersymmetry. This is more difficult as there is no easy notion of coarse-graining (unlike for the fermionic description of LLM), but presumably can still be done. 

We were also able to estimate the complexity of the CFT side by studying the Shannon and participation entropy of states -- but these findings are not easy to relate to the bulk chaos. One could question the relevance of these weak-coupling calculations, however they do tell us at least one important point: at weak coupling, the complexity of the CFT state is not directly related to the the bulk chaos -- the superstratum yields $\log\sqrt{N}$ scaling instead of $\log N$ for the supertube, even though superstratum exhibits stronger wave chaos. This also shows that our bulk chaos hierarchy is not the same thing as the BPS chaos of \cite{Chen:2024oqv,Chen:2026vml}. Rather, the two approaches are complementary and tell us about black-holishness in different coupling regimes. The clear Berry hierarchy on the supergravity side becomes just cherry-picking (or ``Berry-picking`` to say so) on the CFT side. We clearly need to work more to solve the difficult problem of matching the two descriptions of chaos. This is also suggested by the results of \cite{Giataganas:2013dha}, where the authors study the TsT deformations of the bulk corresponding to complex $\beta$-deformations in the CFT: the deformation breaks integrability (though the probes in that paper are very different from ours, strings and not fields) and introduces chaos but there is no direct relation between the details of the deformed CFT and the degree of chaos in the bulk. 

It is useful to contrast our findings with the arguments that appear in the gravitational computations of LMRS \cite{Lin:2022zxd}. In their computation of the spectral form factor (SFF) it was noticed that a certain sum of correlation functions will behave as a Gaussian random matrix. The crucial part of their computation is the Schwarzian dynamics, entering via a gravitational path integral. Since we encounter a similar, long but finite, AdS${}_2$ throat in the superstratum geometry, many conclusions should hold in our case, too. However, in such a computation complex Euclidean saddles will appear, and the difficulty of dealing with the Euclidean description of microstate geometries makes it hard for us to compute SFF. Still, based on the existence of the long and well-defined throat, it seems plausible that such superstrata might exhibit a linear ramp. This expectation challenges the narrative found in the literature that strong chaos should be associated exclusively with fortuitous BPS states \cite{Johnson:2026plw}. This idea was also challenged in \cite{Chen:2025sum, Tierz:2026vrs}, by noting that fortuitous states can exist in an otherwise integrable model. 

Besides, one should bear in mind that fortuity is not a fixed property but can be generated dynamically: recent work \cite{Choi:2025pqr} shows that in a mass-deformed $\mathcal{N}=4$ super-Yang-Mills theory flowing to the Leigh-Strassler fixed point, operators that are monotone in the undeformed $\mathcal{N}=4$ theory become fortuitous in the IR CFT under the relevant deformation. This raises the possibility that the geometric hierarchy we study here — varying supersymmetry and throat length within a family of backgrounds — could likewise be understood as driving a monotone-to-fortuitous transition.

In future work, this scenario can also be tested within the D1-D5 CFT. In \cite{Giusto:2026rpl} the authors study 1/4-BPS states dressed on superstrata-type ``graviton gas`` backgrounds away from the free orbifold point, and compute a nonvanishing three-point coupling connecting two monotone states to a fortuitous one -- direct evidence that the monotone and fortuitous sectors mix once one leaves the free point, rather than staying dynamically decoupled. Together with \cite{Choi:2025pqr}, this suggests that monotone-to-fortuitous transitions may be a fairly generic feature of holographic CFTs in the interacting regime, and consequently also in the supergravity regime.

Finally, we also hope to understand better the relation of our findings to the Berry curvature criterion of BPS chaos as studied in the recent work \cite{Chen:2026vml}. That work deals with very similar problems as ours, but exclusively from the CFT viewpoint, and it is an important task to understand how exactly the two approaches connect.

\section*{Acknowledgments}

We are grateful to Jan Becker, David Berenstein, Nikolay Bobev, Vasil Dimitrov, Raphaël Dulac, Dimitrios Giatagannas, Seunggyu Kim, Ohad Mamroud, Jesse van Muiden, Chiara Toldo, Juan-Diego Urbina, Yoav Zigdon for stimulating discussions. V.~Đ. is especially grateful to Yiming Chen for his lectures and discussions at the CERN Winter School on Supergravity, Strings and Gauge Theory 2026, which inspired this work, and also thanks the organizers of the school. We acknowledge support by the Delta ITP consortium, a program of the Netherlands Organization for Scientific Research (NWO) funded by the Dutch Ministry of Education, Culture and Science (OCW). Work at the Institute of Physics is funded by the Ministry of Education, Science and Technological Development and by the Science Fund of the Republic of Serbia. The work on Sections 3-5 was supported by Russian Science Foundation Grant No. 24-72-10061 [https://rscf.ru/project/24-72-10061/] and performed at Steklov Mathematical Institute of Russian Academy of Sciences (M.\v{C}.).

\bibliography{LLM.bib}

@article{Lin:2004nb,
    author = "Lin, Hai and Lunin, Oleg and Maldacena, Juan Martin",
    title = "{Bubbling AdS space and 1/2 BPS geometries}",
    eprint = "hep-th/0409174",
    archivePrefix = "arXiv",
    reportNumber = "PUPT-2136",
    doi = "10.1088/1126-6708/2004/10/025",
    journal = "JHEP",
    volume = "10",
    pages = "025",
    year = "2004"
}

@article{Berenstein:2004kk,
    author = "Berenstein, David",
    title = "{A Toy model for the AdS / CFT correspondence}",
    eprint = "hep-th/0403110",
    archivePrefix = "arXiv",
    doi = "10.1088/1126-6708/2004/07/018",
    journal = "JHEP",
    volume = "07",
    pages = "018",
    year = "2004"
}

@article{Bena:2004jw,
    author = "Bena, Iosif and Warner, Nicholas P.",
    title = "{A Harmonic family of dielectric flow solutions with maximal supersymmetry}",
    eprint = "hep-th/0406145",
    archivePrefix = "arXiv",
    reportNumber = "USC-04-03, UCLA-04-TEP-10",
    doi = "10.1088/1126-6708/2004/12/021",
    journal = "JHEP",
    volume = "12",
    pages = "021",
    year = "2004"
}

@article{Berenstein:2023vtd,
    author = "Berenstein, David and Maderazo, Elliot and Mancilla, Robinson and Ramirez, Anayeli",
    title = "{Chaotic LLM billiards}",
    eprint = "2305.19321",
    archivePrefix = "arXiv",
    primaryClass = "hep-th",
    doi = "10.1007/JHEP08(2024)056",
    journal = "JHEP",
    volume = "08",
    pages = "056",
    year = "2024"
}

@article{Bena:2020yii,
    author = "Bena, Iosif and Eperon, Felicity and Heidmann, Pierre and Warner, Nicholas P.",
    title = "{The Great Escape: Tunneling out of Microstate Geometries}",
    eprint = "2005.11323",
    archivePrefix = "arXiv",
    primaryClass = "hep-th",
    doi = "10.1007/JHEP04(2021)112",
    journal = "JHEP",
    volume = "04",
    pages = "112",
    year = "2021"
}

@article{Chervonyi:2013eja,
    author = "Chervonyi, Yuri and Lunin, Oleg",
    title = "{(Non)-Integrability of Geodesics in D-brane Backgrounds}",
    eprint = "1311.1521",
    archivePrefix = "arXiv",
    primaryClass = "hep-th",
    doi = "10.1007/JHEP02(2014)061",
    journal = "JHEP",
    volume = "02",
    pages = "061",
    year = "2014"
}

@article{Bena:2017upb,
    author = "Bena, Iosif and Turton, David and Walker, Robert and Warner, Nicholas P.",
    title = "{Integrability and Black-Hole Microstate Geometries}",
    eprint = "1709.01107",
    archivePrefix = "arXiv",
    primaryClass = "hep-th",
    reportNumber = "IPHT-T17-134, IPHT-T17/134",
    doi = "10.1007/JHEP11(2017)021",
    journal = "JHEP",
    volume = "11",
    pages = "021",
    year = "2017"
}

@article{Giataganas:2024hil,
    author = "Giataganas, D. and Kehagias, A. and Riotto, A.",
    title = "{Quasinormal Modes and Universality of the Penrose Limit of Black Hole Photon Rings}",
    eprint = "2403.10605",
    archivePrefix = "arXiv",
    primaryClass = "gr-qc",
    month = "3",
    year = "2024"
}

@article{Frolov:2017kze,
    author = "Frolov, Valeri P. and Krtous, Pavel and Kubiznak, David",
    title = "{Black holes, hidden symmetries, and complete integrability}",
    eprint = "1705.05482",
    archivePrefix = "arXiv",
    primaryClass = "gr-qc",
    doi = "10.1007/s41114-017-0009-9",
    journal = "Living Rev. Rel.",
    volume = "20",
    number = "1",
    pages = "6",
    year = "2017"
}

@article{Bianchi:2020des,
    author = "Bianchi, Massimo and Grillo, Alfredo and Morales, Jose Francisco",
    title = "{Chaos at the rim of black hole and fuzzball shadows}",
    eprint = "2002.05574",
    archivePrefix = "arXiv",
    primaryClass = "hep-th",
    doi = "10.1007/JHEP05(2020)078",
    journal = "JHEP",
    volume = "05",
    pages = "078",
    year = "2020"
}

@article{Warner:2019jll,
    author = "Warner, Nicholas P.",
    title = "{Lectures on Microstate Geometries}",
    eprint = "1912.13108",
    archivePrefix = "arXiv",
    primaryClass = "hep-th",
    month = "12",
    year = "2019"
}

@article{Strominger:1996sh,
    author = "Strominger, Andrew and Vafa, Cumrun",
    title = "{Microscopic origin of the Bekenstein-Hawking entropy}",
    eprint = "hep-th/9601029",
    archivePrefix = "arXiv",
    reportNumber = "HUTP-96-A002, RU-96-01",
    doi = "10.1016/0370-2693(96)00345-0",
    journal = "Phys. Lett. B",
    volume = "379",
    pages = "99--104",
    year = "1996"
}

@article{Maldacena:1997de,
    author = "Maldacena, Juan Martin and Strominger, Andrew and Witten, Edward",
    title = "{Black hole entropy in M theory}",
    eprint = "hep-th/9711053",
    archivePrefix = "arXiv",
    doi = "10.1088/1126-6708/1997/12/002",
    journal = "JHEP",
    volume = "12",
    pages = "002",
    year = "1997"
}

@article{Dijkgraaf:1996it,
    author = "Dijkgraaf, Robbert and Verlinde, Erik P. and Verlinde, Herman L.",
    title = "{Counting dyons in N=4 string theory}",
    eprint = "hep-th/9607026",
    archivePrefix = "arXiv",
    reportNumber = "CERN-TH-96-170",
    doi = "10.1016/S0550-3213(96)00640-2",
    journal = "Nucl. Phys. B",
    volume = "484",
    pages = "543--561",
    year = "1997"
}

@article{Bena:2016agb,
    author = "Bena, Iosif and Martinec, Emil and Turton, David and Warner, Nicholas P.",
    title = "{Momentum Fractionation on Superstrata}",
    eprint = "1601.05805",
    archivePrefix = "arXiv",
    primaryClass = "hep-th",
    reportNumber = "IPHT-T16-004",
    doi = "10.1007/JHEP05(2016)064",
    journal = "JHEP",
    volume = "05",
    pages = "064",
    year = "2016"
}

@article{Bena:2017xbt,
    author = "Bena, Iosif and Giusto, Stefano and Martinec, Emil J. and Russo, Rodolfo and Shigemori, Masaki and Turton, David and Warner, Nicholas P.",
    title = "{Asymptotically-flat supergravity solutions deep inside the black-hole regime}",
    eprint = "1711.10474",
    archivePrefix = "arXiv",
    primaryClass = "hep-th",
    reportNumber = "IPHT-T17-135, QMUL-PH-17-26, YITP-17-127",
    doi = "10.1007/JHEP02(2018)014",
    journal = "JHEP",
    volume = "02",
    pages = "014",
    year = "2018"
}

@article{Bena:2022ldq,
    author = "Bena, Iosif and Martinec, Emil J. and Mathur, Samir D. and Warner, Nicholas P.",
    title = "{Snowmass White Paper: Micro- and Macro-Structure of Black Holes}",
    eprint = "2203.04981",
    archivePrefix = "arXiv",
    primaryClass = "hep-th",
    month = "3",
    year = "2022"
}

@article{Myers:2001aq,
    author = "Myers, Robert C. and Tafjord, Oyvind",
    title = "{Superstars and giant gravitons}",
    eprint = "hep-th/0109127",
    archivePrefix = "arXiv",
    doi = "10.1088/1126-6708/2001/11/009",
    journal = "JHEP",
    volume = "11",
    pages = "009",
    year = "2001"
}

@inproceedings{Berry::LesHouches,
    author = "M.V. Berry",
    title = "Semiclassical Mechanics of Regular and Irregular Motion in Chaotic Behavior of Deterministic Systems",
    booktitle = "Les Houches Lectures XXXVI, G.Iooss, R.H.G.Helleman and R.Stora (eds.) : Amsterdam",
    year = "pp 171-271"
}

@article{Berry:1986lxu,
    author = "Berry, M. V.",
    editor = "Seligman, T. H. and Nishioka, H.",
    title = "{Riemann's Zeta function: A model for quantum chaos?}",
    doi = "10.1007/3-540-17171-1_1",
    journal = "Lect. Notes Phys.",
    volume = "263",
    pages = "1--17",
    year = "1986"
}

@article{Berry:1981mom,
    author = "Berry, M. V.",
    title = "{Quantizing a classically ergodic system: Sinai's billiard and the KKR method}",
    doi = "10.1016/0003-4916(81)90189-5",
    journal = "Annals Phys.",
    volume = "131",
    pages = "163--216",
    year = "1981"
}

@article{Chen:2024oqv,
    author = "Chen, Yiming and Lin, Henry W. and Shenker, Stephen H.",
    title = "{BPS chaos}",
    eprint = "2407.19387",
    archivePrefix = "arXiv",
    primaryClass = "hep-th",
    doi = "10.21468/SciPostPhys.18.2.072",
    journal = "SciPost Phys.",
    volume = "18",
    number = "2",
    pages = "072",
    year = "2025"
}

@article{Chang:2024zqi,
    author = "Chang, Chi-Ming and Lin, Ying-Hsuan",
    title = "{Holographic covering and the fortuity of black holes}",
    eprint = "2402.10129",
    archivePrefix = "arXiv",
    primaryClass = "hep-th",
    month = "2",
    year = "2024"
}

@article{Anempodistov:2025maj,
    author = "Anempodistov, Prokopii and Holguin, Adolfo and Kazakov, Vladimir and Murali, Harish",
    title = "{(Un)solvable Matrix Models for BPS Correlators}",
    eprint = "2508.20094",
    archivePrefix = "arXiv",
    primaryClass = "hep-th",
    month = "8",
    year = "2025"
}

@article{Kazakov:2024ald,
    author = "Kazakov, Vladimir and Murali, Harish and Vieira, Pedro",
    title = "{Huge BPS Operators and Fluid Dynamics in $\mathcal{N}=4$ SYM}",
    eprint = "2406.01798",
    archivePrefix = "arXiv",
    primaryClass = "hep-th",
    month = "6",
    year = "2024"
}

@article{Takayama:2005yq,
    author = "Takayama, Yastoshi and Tsuchiya, Asato",
    title = "{Complex matrix model and fermion phase space for bubbling AdS geometries}",
    eprint = "hep-th/0507070",
    archivePrefix = "arXiv",
    reportNumber = "OU-HET-535",
    doi = "10.1088/1126-6708/2005/10/004",
    journal = "JHEP",
    volume = "10",
    pages = "004",
    year = "2005"
}

@article{Balasubramanian:2005mg,
    author = "Balasubramanian, Vijay and de Boer, Jan and Jejjala, Vishnu and Simon, Joan",
    title = "{The Library of Babel: On the origin of gravitational thermodynamics}",
    eprint = "hep-th/0508023",
    archivePrefix = "arXiv",
    reportNumber = "UPR-1127-7, ITFA-2005-37, DCTP-05-33",
    doi = "10.1088/1126-6708/2005/12/006",
    journal = "JHEP",
    volume = "12",
    pages = "006",
    year = "2005"
}

@article{Mandal:2005wv,
    author = "Mandal, Gautam",
    title = "{Fermions from half-BPS supergravity}",
    eprint = "hep-th/0502104",
    archivePrefix = "arXiv",
    doi = "10.1088/1126-6708/2005/08/052",
    journal = "JHEP",
    volume = "08",
    pages = "052",
    year = "2005"
}

@article{Chen:2025sum,
    author = "Chen, Yiming",
    title = "{Fortuity with a single matrix}",
    eprint = "2511.00790",
    archivePrefix = "arXiv",
    primaryClass = "hep-th",
    month = "11",
    year = "2025"
}

@article{Chen:2026vml,
    author = "Chen, Yiming and Colin-Ellerin, Sean and Mamroud, Ohad and Papadodimas, Kyriakos",
    title = "{Chaos of Berry curvature for BPS microstates}",
    eprint = "2604.23287",
    archivePrefix = "arXiv",
    primaryClass = "hep-th",
    month = "4",
    year = "2026"
}

@article{Berry:1977,
doi = {10.1088/0305-4470/10/12/016},
url = {https://doi.org/10.1088/0305-4470/10/12/016},
year = {1977},
month = {dec},
publisher = {},
volume = {10},
number = {12},
pages = {2083},
author = {M V Berry},
title = {Regular and irregular semiclassical wavefunctions},
journal = {Journal of Physics A: Mathematical and General}
}

@article{Srednicki:1998,
  title = {Correlations in Chaotic Eigenfunctions at Large Separation},
  author = {Hortikar, Sanjay and Srednicki, Mark},
  journal = {Phys. Rev. Lett.},
  volume = {80},
  issue = {8},
  pages = {1646--1649},
  numpages = {0},
  year = {1998},
  month = {Feb},
  publisher = {American Physical Society},
  doi = {10.1103/PhysRevLett.80.1646},
  url = {https://link.aps.org/doi/10.1103/PhysRevLett.80.1646}
}

@article{Urbina:2007fuq,
    author = "Urbina, Juan Diego and Richter, Klaus",
    title = "{Random Wave Functions with boundary and normalization constraints: Quantum statistical physics meets quantum chaos}",
    eprint = "0801.1197",
    archivePrefix = "arXiv",
    primaryClass = "nlin.CD",
    journal = "Eur. Phys. J. ST",
    volume = "145",
    pages = "255--269",
    year = "2007"
}

@article{Sekino:2008he,
    author = "Sekino, Yasuhiro and Susskind, Leonard",
    title = "{Fast Scramblers}",
    eprint = "0808.2096",
    archivePrefix = "arXiv",
    primaryClass = "hep-th",
    reportNumber = "SU-ITP-08-18, OIQP-08-08, SU-ITP-08/18, OIQP-08-08",
    doi = "10.1088/1126-6708/2008/10/065",
    journal = "JHEP",
    volume = "10",
    pages = "065",
    year = "2008"
}

@article{Shenker:2014cwa,
    author = "Shenker, Stephen H. and Stanford, Douglas",
    title = "{Stringy effects in scrambling}",
    eprint = "1412.6087",
    archivePrefix = "arXiv",
    primaryClass = "hep-th",
    doi = "10.1007/JHEP05(2015)132",
    journal = "JHEP",
    volume = "05",
    pages = "132",
    year = "2015"
}

@article{Maldacena:2015waa,
    author = "Maldacena, Juan and Shenker, Stephen H. and Stanford, Douglas",
    title = "{A bound on chaos}",
    eprint = "1503.01409",
    archivePrefix = "arXiv",
    primaryClass = "hep-th",
    doi = "10.1007/JHEP08(2016)106",
    journal = "JHEP",
    volume = "08",
    pages = "106",
    year = "2016"
}

@article{Berenstein:2025ese,
    author = "Berenstein, David and {\v{C}}ubrovi{\'c}, Mihailo and Djuki{\'c}, Vladan",
    title = "{Trapping, chaos and averaging in bubbling AdS spaces}",
    eprint = "2508.09669",
    archivePrefix = "arXiv",
    primaryClass = "hep-th",
    doi = "10.1007/JHEP02(2026)157",
    journal = "JHEP",
    volume = "02",
    pages = "157",
    year = "2026"
}

@article{Giusto:2019qig,
    author = "Giusto, Stefano and Rawash, Sami and Turton, David",
    title = "{AdS$_{3}$ holography at dimension two}",
    eprint = "1904.12880",
    archivePrefix = "arXiv",
    primaryClass = "hep-th",
    doi = "10.1007/JHEP07(2019)171",
    journal = "JHEP",
    volume = "07",
    pages = "171",
    year = "2019"
}

@article{Giusto:2015dfa,
    author = "Giusto, Stefano and Moscato, Emanuele and Russo, Rodolfo",
    title = "{AdS$_{3}$ holography for 1/4 and 1/8 BPS geometries}",
    eprint = "1507.00945",
    archivePrefix = "arXiv",
    primaryClass = "hep-th",
    reportNumber = "DFPD-15-TH-15, QMUL-PH-15-12",
    doi = "10.1007/JHEP11(2015)004",
    journal = "JHEP",
    volume = "11",
    pages = "004",
    year = "2015"
}

@article{Bena:2013pda,
    author = "Bena, Iosif and El-Showk, Sheer and Vercnocke, Bert",
    editor = "Bellucci, Stefano",
    title = "{Black Holes in String Theory}",
    doi = "10.1007/978-3-319-00215-6_2",
    journal = "Springer Proc. Phys.",
    volume = "144",
    pages = "59--178",
    year = "2013"
}

@article{Bena:2025pcy,
    author = "Bena, Iosif and Warner, Nicholas P.",
    title = "{Microstate Geometries}",
    eprint = "2503.17310",
    archivePrefix = "arXiv",
    primaryClass = "hep-th",
    month = "3",
    year = "2025"
}

@article{Lin:2022zxd,
    author = "Lin, Henry W. and Maldacena, Juan and Rozenberg, Liza and Shan, Jieru",
    title = "{Looking at supersymmetric black holes for a very long time}",
    eprint = "2207.00408",
    archivePrefix = "arXiv",
    primaryClass = "hep-th",
    doi = "10.21468/SciPostPhys.14.5.128",
    journal = "SciPost Phys.",
    volume = "14",
    number = "5",
    pages = "128",
    year = "2023"
}

@article{Lin:2022rzw,
    author = "Lin, Henry W. and Maldacena, Juan and Rozenberg, Liza and Shan, Jieru",
    title = "{Holography for people with no time}",
    eprint = "2207.00407",
    archivePrefix = "arXiv",
    primaryClass = "hep-th",
    doi = "10.21468/SciPostPhys.14.6.150",
    journal = "SciPost Phys.",
    volume = "14",
    number = "6",
    pages = "150",
    year = "2023"
}

@article{Johnson:2026plw,
    author = "Johnson, Clifford V.",
    title = "{Fortuitous Chaos, BPS Black Holes, and Random Matrices}",
    eprint = "2601.17122",
    archivePrefix = "arXiv",
    primaryClass = "hep-th",
    month = "1",
    year = "2026"
}

@article{Shigemori:2019orj,
    author = "Shigemori, Masaki",
    title = "{Counting Superstrata}",
    eprint = "1907.03878",
    archivePrefix = "arXiv",
    primaryClass = "hep-th",
    reportNumber = "YITP-19-61",
    doi = "10.1007/JHEP10(2019)017",
    journal = "JHEP",
    volume = "10",
    pages = "017",
    year = "2019"
}

@article{Mayerson:2020acj,
    author = "Mayerson, Daniel R. and Shigemori, Masaki",
    title = "{Counting D1-D5-P microstates in supergravity}",
    eprint = "2010.04172",
    archivePrefix = "arXiv",
    primaryClass = "hep-th",
    doi = "10.21468/SciPostPhys.10.1.018",
    journal = "SciPost Phys.",
    volume = "10",
    number = "1",
    pages = "018",
    year = "2021"
}

@article{Hughes:2025tdy,
    author = "Hughes, Marcel R. R. and Shigemori, Masaki",
    title = "{Fortuity and supergravity}",
    eprint = "2505.14888",
    archivePrefix = "arXiv",
    primaryClass = "hep-th",
    reportNumber = "YITP-25-74",
    doi = "10.1007/JHEP03(2026)130",
    journal = "JHEP",
    volume = "03",
    pages = "130",
    year = "2026"
}

@article{Bena:2018bbd,
    author = "Bena, Iosif and Heidmann, Pierre and Turton, David",
    title = "{AdS$_{2}$ holography: mind the cap}",
    eprint = "1806.02834",
    archivePrefix = "arXiv",
    primaryClass = "hep-th",
    doi = "10.1007/JHEP12(2018)028",
    journal = "JHEP",
    volume = "12",
    pages = "028",
    year = "2018"
}

@article{Dabholkar:2004yr,
    author = "Dabholkar, Atish",
    title = "{Exact counting of black hole microstates}",
    eprint = "hep-th/0409148",
    archivePrefix = "arXiv",
    reportNumber = "SLAC-PUB-10717, SU-ITP-04-36, TIFR-TH-04-23",
    doi = "10.1103/PhysRevLett.94.241301",
    journal = "Phys. Rev. Lett.",
    volume = "94",
    pages = "241301",
    year = "2005"
}

@article{Giusto:2012yz,
    author = "Giusto, Stefano and Lunin, Oleg and Mathur, Samir D. and Turton, David",
    title = "{D1-D5-P microstates at the cap}",
    eprint = "1211.0306",
    archivePrefix = "arXiv",
    primaryClass = "hep-th",
    doi = "10.1007/JHEP02(2013)050",
    journal = "JHEP",
    volume = "02",
    pages = "050",
    year = "2013"
}

@article{Giusto:2011fy,
    author = "Giusto, Stefano and Russo, Rodolfo and Turton, David",
    title = "{New D1-D5-P geometries from string amplitudes}",
    eprint = "1108.6331",
    archivePrefix = "arXiv",
    primaryClass = "hep-th",
    reportNumber = "DFPD-11-TH-14, QMUL-PH-11-12",
    doi = "10.1007/JHEP11(2011)062",
    journal = "JHEP",
    volume = "11",
    pages = "062",
    year = "2011"
}

@article{Lunin:2001fv,
    author = "Lunin, Oleg and Mathur, Samir D.",
    title = "{Metric of the multiply wound rotating string}",
    eprint = "hep-th/0105136",
    archivePrefix = "arXiv",
    reportNumber = "OHSTPY-HEP-T-01-014",
    doi = "10.1016/S0550-3213(01)00321-2",
    journal = "Nucl. Phys. B",
    volume = "610",
    pages = "49--76",
    year = "2001"
}

@article{Lunin:2002iz,
    author = "Lunin, Oleg and Maldacena, Juan Martin and Maoz, Liat",
    title = "{Gravity solutions for the D1-D5 system with angular momentum}",
    eprint = "hep-th/0212210",
    archivePrefix = "arXiv",
    month = "12",
    year = "2002"
}

@article{Raju:2018xue,
    author = "Raju, Suvrat and Shrivastava, Pushkal",
    title = "{Critique of the fuzzball program}",
    eprint = "1804.10616",
    archivePrefix = "arXiv",
    primaryClass = "hep-th",
    doi = "10.1103/PhysRevD.99.066009",
    journal = "Phys. Rev. D",
    volume = "99",
    number = "6",
    pages = "066009",
    year = "2019"
}

@article{Shigemori:2020yuo,
    author = "Shigemori, Masaki",
    title = "{Superstrata}",
    eprint = "2002.01592",
    archivePrefix = "arXiv",
    primaryClass = "hep-th",
    reportNumber = "YITP-20-16",
    doi = "10.1007/s10714-020-02698-8",
    journal = "Gen. Rel. Grav.",
    volume = "52",
    number = "5",
    pages = "51",
    year = "2020"
}

@article{Giusto:2026rpl,
    author = "Giusto, Stefano and Inglis, James and Russo, Rodolfo",
    title = "{Fortuity beyond counting: an explicit construction}",
    eprint = "2606.20353",
    archivePrefix = "arXiv",
    primaryClass = "hep-th",
    month = "6",
    year = "2026"
}

@article{Tierz:2026vrs,
    author = "Tierz, Miguel",
    title = "{BPS spectra of $\operatorname{tr}[\Psi^p]$ matrix models for odd $p$}",
    eprint = "2604.27164",
    archivePrefix = "arXiv",
    primaryClass = "hep-th",
    month = "4",
    year = "2026"
}

@article{Mirlin:2000,
  title={Statistics of energy levels and eigenfunctions in disordered systems},
  author={Mirlin, Alexander D},
  journal={Physics Reports},
  volume={326},
  number={4-6},
  pages={259--382},
  year={2000},
  publisher={Elsevier}
}

@article{Mirlin:2002,
  title = {Wave function correlations on the ballistic scale: Exploring quantum chaos by quantum disorder},
  author = {Gornyi, I. V. and Mirlin, A. D.},
  journal = {Phys. Rev. E},
  volume = {65},
  issue = {2},
  pages = {025202(R)},
  numpages = {4},
  year = {2002},
  month = {Jan},
  publisher = {American Physical Society},
  doi = {10.1103/PhysRevE.65.025202},
  url = {https://link.aps.org/doi/10.1103/PhysRevE.65.025202}
}

@article{Urbina:2003xip,
    author = "Urbina, Juan Diego and Richter, Klaus",
    title = "{Semiclassical Construction of Random Wave Functions for Confined Systems}",
    eprint = "nlin/0309004",
    archivePrefix = "arXiv",
    doi = "10.1103/PhysRevE.70.015201",
    month = "9",
    year = "2003"
}

@article{Balasubramanian:2018yjq,
    author = "Balasubramanian, Vijay and Berenstein, David and Lewkowycz, Aitor and Miller, Alexandra and Parrikar, Onkar and Rabideau, Charles",
    title = "{Emergent classical spacetime from microstates of an incipient black hole}",
    eprint = "1810.13440",
    archivePrefix = "arXiv",
    primaryClass = "hep-th",
    doi = "10.1007/JHEP01(2019)197",
    journal = "JHEP",
    volume = "01",
    pages = "197",
    year = "2019"
}

@article{Berenstein:2022srd,
    author = "Berenstein, David and Wang, Shannon",
    title = "{BPS coherent states and localization}",
    eprint = "2203.15820",
    archivePrefix = "arXiv",
    primaryClass = "hep-th",
    doi = "10.1007/JHEP08(2022)164",
    journal = "JHEP",
    volume = "08",
    pages = "164",
    year = "2022"
}

@article{Lin:2022wdr,
    author = "Lin, Hai",
    title = "{Coherent state excitations and string-added coherent states in gauge-gravity correspondence}",
    eprint = "2206.06524",
    archivePrefix = "arXiv",
    primaryClass = "hep-th",
    doi = "10.1016/j.nuclphysb.2022.116066",
    journal = "Nucl. Phys. B",
    volume = "986",
    pages = "116066",
    year = "2023"
}

@book{Haake:book,
author = {Haake, Fritz},
title = {Quantum Signatures of Chaos},
year = {2006},
isbn = {3540677232},
publisher = {Springer-Verlag},
address = {Berlin, Heidelberg}
}

@book{Stockmann:book,
  title={Quantum Chaos: An Introduction},
  author={St{\"o}ckmann, Hans-J{\"u}rgen},
  year={1999},
  publisher={Cambridge University Press},
  address={Cambridge},
  isbn={9780521027151}
}

@article{Almheiri:2020cfm,
    author = "Almheiri, Ahmed and Hartman, Thomas and Maldacena, Juan and Shaghoulian, Edgar and Tajdini, Amirhossein",
    title = "{The entropy of Hawking radiation}",
    eprint = "2006.06872",
    archivePrefix = "arXiv",
    primaryClass = "hep-th",
    doi = "10.1103/RevModPhys.93.035002",
    journal = "Rev. Mod. Phys.",
    volume = "93",
    number = "3",
    pages = "035002",
    year = "2021"
}

@article{Giataganas:2026ctn,
    author = "Giataganas, D. and Giudice, G. F. and Kehagias, A. and Quevedo, F. and Riotto, A.",
    title = "{Black Hole Photon Rings Saturate the Quantum Chaos Bound}",
    eprint = "2605.29923",
    archivePrefix = "arXiv",
    primaryClass = "hep-th",
    reportNumber = "CERN-TH-2026-100",
    month = "5",
    year = "2026"
}

@article{Shenker:2013pqa,
    author = "Shenker, Stephen H. and Stanford, Douglas",
    title = "{Black holes and the butterfly effect}",
    eprint = "1306.0622",
    archivePrefix = "arXiv",
    primaryClass = "hep-th",
    reportNumber = "SU-ITP-13-08",
    doi = "10.1007/JHEP03(2014)067",
    journal = "JHEP",
    volume = "03",
    pages = "067",
    year = "2014"
}

@article{Polchinski:2015cea,
    author = "Polchinski, Joseph",
    title = "{Chaos in the black hole S-matrix}",
    eprint = "1505.08108",
    archivePrefix = "arXiv",
    primaryClass = "hep-th",
    month = "5",
    year = "2015"
}

@article{Balasubramanian:2022gmo,
    author = "Balasubramanian, Vijay and Lawrence, Albion and Magan, Javier M. and Sasieta, Martin",
    title = "{Microscopic Origin of the Entropy of Black Holes in General Relativity}",
    eprint = "2212.02447",
    archivePrefix = "arXiv",
    primaryClass = "hep-th",
    doi = "10.1103/PhysRevX.14.011024",
    journal = "Phys. Rev. X",
    volume = "14",
    number = "1",
    pages = "011024",
    year = "2024"
}

@article{Cotler:2016fpe,
    author = "Cotler, Jordan S. and Gur-Ari, Guy and Hanada, Masanori and Polchinski, Joseph and Saad, Phil and Shenker, Stephen H. and Stanford, Douglas and Streicher, Alexandre and Tezuka, Masaki",
    title = "{Black Holes and Random Matrices}",
    eprint = "1611.04650",
    archivePrefix = "arXiv",
    primaryClass = "hep-th",
    reportNumber = "SU-ITP-16-19, SU-ITP-16/19, YITP-16-124",
    doi = "10.1007/JHEP05(2017)118",
    journal = "JHEP",
    volume = "05",
    pages = "118",
    year = "2017",
    note = "[Erratum: JHEP 09, 002 (2018)]"
}

@article{Saad:2018bqo,
    author = "Saad, Phil and Shenker, Stephen H. and Stanford, Douglas",
    title = "{A semiclassical ramp in SYK and in gravity}",
    eprint = "1806.06840",
    archivePrefix = "arXiv",
    primaryClass = "hep-th",
    month = "6",
    year = "2018"
}

@article{Saad:2019lba,
    author = "Saad, Phil and Shenker, Stephen H. and Stanford, Douglas",
    title = "{JT gravity as a matrix integral}",
    eprint = "1903.11115",
    archivePrefix = "arXiv",
    primaryClass = "hep-th",
    month = "3",
    year = "2019"
}

@article{Basu:2011di,
    author = "Basu, Pallab and Pando Zayas, Leopoldo A.",
    title = "{Chaos rules out integrability of strings on AdS$_5 \times T^{1,1}$}",
    eprint = "1103.4107",
    archivePrefix = "arXiv",
    primaryClass = "hep-th",
    reportNumber = "MCTP-11-12",
    doi = "10.1016/j.physletb.2011.04.063",
    journal = "Phys. Lett. B",
    volume = "700",
    pages = "243--248",
    year = "2011"
}

@article{Djukic:2023dgk,
    author = "Djuki{\'c}, Vladan and {\v{C}}ubrovi{\'c}, Mihailo",
    title = "{Correlation functions for open strings and chaos}",
    eprint = "2310.15697",
    archivePrefix = "arXiv",
    primaryClass = "hep-th",
    doi = "10.1007/JHEP04(2024)025",
    journal = "JHEP",
    volume = "04",
    pages = "025",
    year = "2024"
}

@article{Giataganas:2021ghs,
    author = "Giataganas, Dimitrios",
    title = "{Chaotic Motion near Black Hole and Cosmological Horizons}",
    eprint = "2112.02081",
    archivePrefix = "arXiv",
    primaryClass = "hep-th",
    doi = "10.1002/prop.202200001",
    journal = "Fortsch. Phys.",
    volume = "70",
    number = "1",
    pages = "2200001",
    year = "2022"
}

@article{Santos:2019,
  title = {Self-averaging in many-body quantum systems out of equilibrium: Chaotic systems},
  author = {Schiulaz, Mauro and Torres-Herrera, E. Jonathan and P\'erez-Bernal, Francisco and Santos, Lea F.},
  journal = {Phys. Rev. B},
  volume = {101},
  issue = {17},
  pages = {174312},
  numpages = {16},
  year = {2020},
  month = {May},
  publisher = {American Physical Society},
  doi = {10.1103/PhysRevB.101.174312},
  url = {https://link.aps.org/doi/10.1103/PhysRevB.101.174312}
}

@article{Santos:2011,
  title = {Entropy of Isolated Quantum Systems after a Quench},
  author = {Santos, Lea F. and Polkovnikov, Anatoli and Rigol, Marcos},
  journal = {Phys. Rev. Lett.},
  volume = {107},
  issue = {4},
  pages = {040601},
  numpages = {4},
  year = {2011},
  month = {Jul},
  publisher = {American Physical Society},
  doi = {10.1103/PhysRevLett.107.040601},
  url = {https://link.aps.org/doi/10.1103/PhysRevLett.107.040601}
}

@article{Choi:2025pqr,
    author = "Choi, Jaehyeok and Kim, Seunggyu",
    title = "{Fortuity and relevant deformation}",
    eprint = "2512.12674",
    archivePrefix = "arXiv",
    primaryClass = "hep-th",
    doi = "10.1007/JHEP07(2026)097",
    journal = "JHEP",
    volume = "07",
    pages = "097",
    year = "2026"
}

@article{Giataganas:2013dha,
    author = "Giataganas, Dimitrios and Pando Zayas, Leopoldo A. and Zoubos, Konstantinos",
    title = "{On Marginal Deformations and Non-Integrability}",
    eprint = "1311.3241",
    archivePrefix = "arXiv",
    primaryClass = "hep-th",
    reportNumber = "MCTP-13-34",
    doi = "10.1007/JHEP01(2014)129",
    journal = "JHEP",
    volume = "01",
    pages = "129",
    year = "2014"
}

\end{document}